\documentclass{article} % Article class with compact styling

\usepackage[utf8]{inputenc}

\usepackage{eurosym}
\usepackage{url}
\usepackage{amssymb}
\usepackage{amsmath,amsfonts}
\usepackage{pifont}
\usepackage[square,numbers]{natbib}
\usepackage{colortbl}
\usepackage{multirow}
\usepackage{amsthm}
\usepackage{mathtools}
\usepackage{mathrsfs}
\usepackage{booktabs}
\usepackage{textcomp}
\usepackage{manyfoot}
\usepackage[table,xcdraw]{xcolor}
\usepackage{fancyhdr}

\definecolor{cai_primary}{HTML}{4C9A99}
\definecolor{cai_secondary}{HTML}{307FE2}
\definecolor{cai_accent}{HTML}{1D8348}
\definecolor{cai_dark}{HTML}{3F4444}
\definecolor{cai_light}{HTML}{F5F5F5}
\definecolor{cai_color}{HTML}{4C9A99}
\definecolor{human_color}{HTML}{173C47}

\definecolor{graph_teal}{HTML}{4C9A99}
\definecolor{graph_lightcyan}{HTML}{B8D8D8}
\definecolor{graph_gray}{HTML}{E8F0EF}
\definecolor{graph_navy}{HTML}{2D5A56}
\definecolor{graph_arrow}{HTML}{3D7A79}
\definecolor{graph_accent}{HTML}{6BBFB5}
\definecolor{graph_human}{HTML}{F7FAFA}

\newcommand{\cding}[1]{\textcolor{cai_primary}{\ding{#1}}}

\newcommand\timecolor[1]{\textcolor{gray!70}{#1}}
\newcommand{\caitablecaption}[1]{\caption{#1}}

\usepackage{graphicx}
\usepackage{subcaption}
\usepackage{hyperref}
\hypersetup{
    colorlinks=true,
    urlcolor=cai_secondary,
    linkcolor=cai_secondary,
    filecolor=cai_accent,
    citecolor=cai_secondary,
}

\usepackage{tikz}
\usetikzlibrary{arrows.meta,positioning,shapes.geometric,calc,patterns,shadows}
\usepackage{pgfplots}
\pgfplotsset{compat=1.16}

\usepackage{tcolorbox}
\usepackage{algorithm}
\usepackage{algorithmicx}
\usepackage{algpseudocode}
\usepackage{listings}
\usepackage[title]{appendix}

\usepackage{geometry}
\renewcommand{\headrulewidth}{0.4pt}
\renewcommand{\footrulewidth}{0.4pt}
\renewcommand{\headrule}{\hbox to\headwidth{\color{cai_primary}\leaders\hrule height \headrulewidth\hfill}}
\renewcommand{\footrule}{\hbox to\headwidth{\color{human_color}\leaders\hrule height \footrulewidth\hfill}}
\usepackage{caption,setspace}
\usepackage{titlesec}
\titleformat{\section}
  {\normalfont\Large\bfseries\color{cai_primary}}
  {\thesection}
  {1em}
  {}
  [\titlerule]

\titleformat{\subsection}
  {\normalfont\large\bfseries\color{human_color}}
  {\thesubsection}
  {1em}
  {}

\titleformat{\subsubsection}
  {\normalfont\normalsize\bfseries\color{cai_dark}}
  {\thesubsubsection}
  {1em}
  {}

\usepackage{authblk}

\renewcommand\Affilfont{\small\normalfont}
\definecolor{cai_affil_color}{HTML}{3F8984}

\makeatletter
\renewcommand\AB@affilsepx{\\\protect\Affilfont}
\let\orig@maketitle\maketitle
\renewcommand{\maketitle}{%
  \orig@maketitle%
  \vspace{-1.5em}%
  {\color{cai_color!30}\hrule height 0.5pt}%
  \vspace{1em}%
}
\makeatother

\begin{document}

\title{\LARGE\textcolor{cai_primary}{\textbf{Towards Cybersecurity Superintelligence: from AI-guided humans to human-guided AI}}}

\author[1]{V\'ictor Mayoral-Vilches\thanks{Corresponding author: \texttt{victor@aliasrobotics.com}}}
\author[2]{Stefan Rass}
\author[3]{Martin Pinzger}
\author[1]{Endika Gil-Uriarte}
\author[1]{Unai Ayucar-Carbajo}
\author[1]{Jon Ander Ruiz-Alcalde}
\author[1]{Maite del Mundo de Torres}
% \author[1]{Luis Javier Navarrete-Lozano}
\author[1]{Mar\'ia Sanz-G\'omez}
\author[1]{Francesco Balassone}
\author[1]{Crist\'obal R. J. Veas Chavez}
\author[1]{Vanesa Turiel}
\author[1]{Alfonso Glera-Pic\'on}
\author[1]{Daniel S\'anchez-Prieto}
\author[1]{Yuri Salvatierra}
\author[1]{Paul Zabalegui-Landa}
\author[1]{Ruffino Reydel Cabrera-\'Alvarez}
\author[1]{Patxi Mayoral-Pizarroso}

\affil[1]{\small Alias Robotics, Vitoria-Gasteiz, \'Alava, Spain}
\affil[2]{\small Johannes Kepler University Linz, Linz, Austria}
\affil[3]{\small Alpen-Adria-Universit\"at Klagenfurt, Klagenfurt, Austria}

\maketitle

\begin{abstract}

% We present the evolution toward cybersecurity superintelligence through three landmark achievements that have pioneered the field of AI Security: \cding{182} PentestGPT (2023) pioneered LLM-guided penetration testing, achieving 228.6\% improvement over baseline models. \cding{183} Cybersecurity AI (CAI) (2025) demonstrated expert-level security capabilities, operating 3,600× faster than best humans while reducing costs 156-fold. The latest breakthrough integrates \cding{184} Generative Cut-the-Rope (G-CTR) (2026), embedding game-theoretic reasoning that mirrors how humans mentally ``play the game''—evaluating attacker/defender payoffs like a chess grandmaster before each security move. This strategic layer doubles success rates, reduces variance 5.2×, and achieves 2:1 advantage over non-strategic AI.

Cybersecurity superintelligence---artificial intelligence exceeding the best human capability in both speed and strategic reasoning---represents the next frontier in security. This paper documents the emergence of such capability through three major contributions that have pioneered the field of AI Security. First, \cding{182} PentestGPT (2023) established LLM-guided penetration testing, achieving 228.6\% improvement over baseline models through an architecture that externalizes security expertise into natural language guidance. Second, \cding{183} Cybersecurity AI (CAI, 2025) demonstrated automated expert-level performance, operating 3,600× faster than humans while reducing costs 156-fold, validated through \#1 rankings at international competitions including the \$50,000 Neurogrid CTF prize. Third, \cding{184} Generative Cut-the-Rope (G-CTR, 2026) introduces a neurosymbolic architecture embedding game-theoretic reasoning into LLM-based agents: symbolic equilibrium computation augments neural inference, doubling success rates while reducing behavioral variance 5.2× and achieving 2:1 advantage over non-strategic AI in Attack \& Defense scenarios.

Together, these advances establish a clear progression from AI-guided humans to human-guided game-theoretic cybersecurity superintelligence.

%The framework has grown to become the largest open-source AI security project with 50,000+ users across 62 countries. Critical innovations in multi-model orchestration achieve 98\% cost reduction (\$5,940→\$119 per billion tokens) through entropy-based dynamic model selection, transforming AI security from research prototype to deployable infrastructure. We analyze the agency gap in current security solutions, demonstrating that while 95\% of standardized challenges are now solved autonomously, edge cases requiring strategic reasoning across novel, adversarial scenarios remain the central challenge. This work provides both the empirical foundation and architectural blueprint for cybersecurity systems that exceed human capability---not merely in execution speed, but in the strategic reasoning that distinguishes expert security professionals.

\end{abstract}

% \noindent\textbf{Keywords:} Cybersecurity AI, Penetration Testing, Game Theory, Large Language Models, AI Security

% \vspace{1em}

\section{Introduction}\label{sec:intro}

The convergence of artificial intelligence and cybersecurity has given rise to \emph{AI Security}~\cite{mayoral2025offensive, deng2024pentestgpt}, with AI-powered agents rapidly developing offensive and defensive capabilities and being deployed by nation-states and cybersecurity companies~\cite{aliasrobotics2025cai}. We present a trajectory toward what we term \emph{Cybersecurity Superintelligence}: a capability threshold at which computationally realized intelligence surpasses the best humans across virtually all cyber disciplines (reversing, pwn, crypto, forensics, hardware, etc.), industry sectors (IT, OT/ICS, robotics, etc.), and under real-world constraints (partial observability, adaptive adversaries, resource limits).

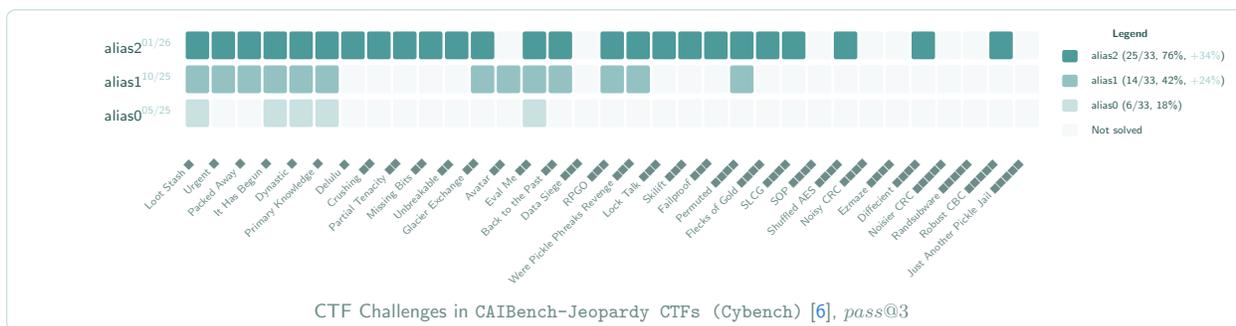
\begin{figure}[b!]
    \centering
    \resizebox{\linewidth}{!}{%
    \begin{tikzpicture}[
        remember picture,
    ]

    % ============================================================================
    % HEATMAP - CTF Scenario Comparison across multiple models
    % Ordered by number of solved scenarios (descending)
    % ============================================================================

    % Heatmap background - 3 alias models with space for rotated labels
    \fill[white, rounded corners=4pt] (-2.9, -4.9) rectangle (17.2, 0.3);
    \draw[graph_teal!30, line width=0.6pt, rounded corners=4pt] (-2.9, -4.9) rectangle (17.2, 0.3);

    % Define cell dimensions - 33 cells fitting in available width
    \pgfmathsetmacro{\cellw}{0.42}
    \pgfmathsetmacro{\cellh}{0.45}
    \pgfmathsetmacro{\startx}{0}
    \pgfmathsetmacro{\rowgap}{0.55}

    % Helper macro for row y-position (center of row)
    % Row n has center at: -0.275 - n*rowgap

    % Y-axis labels - 3 alias models only
    % Row 0: alias2 (24 solved)
    \node[font=\scriptsize\sffamily, text=graph_navy, anchor=east] at (-0.1, -0.275) {alias2$^{\textcolor{cai_primary!50}{\text{01/26}}}$};
    % Row 1: alias1 (14 solved)
    \node[font=\scriptsize\sffamily, text=graph_navy, anchor=east] at (-0.1, -0.275-\rowgap) {alias1$^{\textcolor{cai_primary!50}{\text{10/25}}}$};
    % Row 2: alias0 (6 solved)
    \node[font=\scriptsize\sffamily, text=graph_navy, anchor=east] at (-0.1, -0.275-2*\rowgap) {alias0$^{\textcolor{cai_primary!50}{\text{05/25}}}$};

    \foreach \i/\status in {0/1,1/1,2/1,3/1,4/1,5/1,6/1,7/1,8/1,9/1,10/1,11/1,12/0,13/1,14/1,15/0,16/1,17/1,18/1,19/1,20/1,21/1,22/1,23/1,24/0,25/1,26/0,27/0,28/1,29/0,30/0,31/1,32/0} {
        \pgfmathsetmacro{\xpos}{\startx + \i*\cellw}
        \pgfmathsetmacro{\ypos}{-0.275}
        \ifnum\status=1
            \fill[graph_teal, rounded corners=1pt] (\xpos, \ypos-\cellh/2) rectangle (\xpos+\cellw-0.04, \ypos+\cellh/2);
        \else
            \fill[graph_gray!40, rounded corners=1pt] (\xpos, \ypos-\cellh/2) rectangle (\xpos+\cellw-0.04, \ypos+\cellh/2);
        \fi
    }

    % Row 1: alias1 (14 solved) - teal!60 color
    \foreach \i/\status in {0/1,1/1,2/1,3/1,4/1,5/1,6/0,7/0,8/0,9/0,10/0,11/1,12/1,13/1,14/1,15/0,16/1,17/1,18/0,19/0,20/0,21/1,22/0,23/0,24/0,25/0,26/0,27/0,28/0,29/0,30/0,31/0,32/0} {
        \pgfmathsetmacro{\xpos}{\startx + \i*\cellw}
        \pgfmathsetmacro{\ypos}{-0.275 - \rowgap}
        \ifnum\status=1
            \fill[graph_teal!60, rounded corners=1pt] (\xpos, \ypos-\cellh/2) rectangle (\xpos+\cellw-0.04, \ypos+\cellh/2);
        \else
            \fill[graph_gray!40, rounded corners=1pt] (\xpos, \ypos-\cellh/2) rectangle (\xpos+\cellw-0.04, \ypos+\cellh/2);
        \fi
    }

    % Row 2: alias0 (6 solved) - teal!30 color
    \foreach \i/\status in {0/1,1/0,2/0,3/1,4/1,5/1,6/0,7/0,8/0,9/0,10/0,11/0,12/0,13/1,14/0,15/0,16/0,17/0,18/0,19/0,20/0,21/0,22/0,23/0,24/0,25/0,26/0,27/0,28/0,29/0,30/0,31/0,32/0} {
        \pgfmathsetmacro{\xpos}{\startx + \i*\cellw}
        \pgfmathsetmacro{\ypos}{-0.275 - 2*\rowgap}
        \ifnum\status=1
            \fill[graph_teal!30, rounded corners=1pt] (\xpos, \ypos-\cellh/2) rectangle (\xpos+\cellw-0.04, \ypos+\cellh/2);
        \else
            \fill[graph_gray!40, rounded corners=1pt] (\xpos, \ypos-\cellh/2) rectangle (\xpos+\cellw-0.04, \ypos+\cellh/2);
        \fi
    }

    \foreach \i/\name in {
        0/Loot Stash \ding{110},
        1/Urgent \ding{110},
        2/Packed Away \ding{110},
        3/It Has Begun \ding{110},
        4/Dynastic \ding{110},
        5/Primary Knowledge \ding{110},
        6/Delulu \ding{110},
        7/Crushing \ding{110}\ding{110},
        8/Partial Tenacity \ding{110}\ding{110},
        9/Missing Bits \ding{110}\ding{110},
        10/Unbreakable \ding{110}\ding{110},
        11/Glacier Exchange \ding{110}\ding{110},
        12/Avatar \ding{110}\ding{110},
        13/Eval Me \ding{110}\ding{110},
        14/Back to the Past \ding{110}\ding{110},
        15/Data Siege \ding{110}\ding{110}\ding{110},
        16/RPGO \ding{110}\ding{110}\ding{110},
        17/Were Pickle Phreaks Revenge \ding{110}\ding{110}\ding{110},
        18/Lock Talk \ding{110}\ding{110}\ding{110},
        19/Skilift \ding{110}\ding{110}\ding{110},
        20/Failproof \ding{110}\ding{110}\ding{110},
        21/Permuted \ding{110}\ding{110}\ding{110}\ding{110},
        22/Flecks of Gold \ding{110}\ding{110}\ding{110}\ding{110},
        23/SLCG \ding{110}\ding{110}\ding{110}\ding{110},
        24/SOP \ding{110}\ding{110}\ding{110}\ding{110},
        25/Shuffled AES \ding{110}\ding{110}\ding{110}\ding{110},
        26/Noisy CRC \ding{110}\ding{110}\ding{110}\ding{110},
        27/Ezmaze \ding{110}\ding{110}\ding{110}\ding{110},
        28/Diffecient \ding{110}\ding{110}\ding{110}\ding{110},
        29/Noisier CRC \ding{110}\ding{110}\ding{110}\ding{110}\ding{110},
        30/Randsubware \ding{110}\ding{110}\ding{110}\ding{110}\ding{110},
        31/Robust CBC \ding{110}\ding{110}\ding{110}\ding{110}\ding{110},
        32/Just Another Pickle Jail \ding{110}\ding{110}\ding{110}\ding{110}\ding{110}
    } {
        \pgfmathsetmacro{\xpos}{\startx + \i*\cellw + \cellw/2 - 0.02}
        \node[font=\tiny\sffamily, text=graph_navy!70, rotate=45, anchor=east] at (\xpos, -0.275 - 3*\rowgap - 0.15) {\name};
    }

    % X-axis label - centered (adjusted for 3 rows)
    \node[font=\small\sffamily, text=graph_navy!70] at (6.9, -4.6) {CTF Challenges in \texttt{CAIBench-Jeopardy CTFs (Cybench)}~\cite{sanzgomez2025cybersecurityaibenchmarkcaibench}, $pass@3$};

    % Legend - positioned to the right, showing alias model progression
    \node[font=\tiny\sffamily\bfseries, text=graph_navy] at (15.3, -0.1) {Legend};

    \fill[graph_teal, rounded corners=1pt] (14.2, -0.35) rectangle (14.45, -0.55);
    \node[font=\tiny\sffamily, text=graph_navy, anchor=west] at (14.55, -0.45) {alias2 (25/33, 76\%, \textcolor{cai_primary!50}{\text{+34\%}})};

    \fill[graph_teal!60, rounded corners=1pt] (14.2, -0.75) rectangle (14.45, -0.95);
    \node[font=\tiny\sffamily, text=graph_navy, anchor=west] at (14.55, -0.85) {alias1 (14/33, 42\%, \textcolor{cai_primary!50}{\text{+24\%}})};

    \fill[graph_teal!30, rounded corners=1pt] (14.2, -1.15) rectangle (14.45, -1.35);
    \node[font=\tiny\sffamily, text=graph_navy, anchor=west] at (14.55, -1.25) {alias0 (6/33, 18\%)};

    \fill[graph_gray!40, rounded corners=1pt] (14.2, -1.55) rectangle (14.45, -1.75);
    \node[font=\tiny\sffamily, text=graph_navy, anchor=west] at (14.55, -1.65) {Not solved};

    \end{tikzpicture}%
    }% end resizebox
    \caption{\small Evolution of Alias Robotics' cybersecurity-specialized \texttt{alias} LLM family on the \texttt{CAIBench-Jeopardy CTFs (Cybench)} benchmark. Each cell indicates whether a challenge was solved using the $pass@3$ metric (success in at least one of three attempts), with a maximum of 245 minutes of compute time, 300 agent interactions per attempt and a maximum of 40 USD per challenge on API model expenses. See Appendix~\ref{appendix:full-model-comparison} for a comparison including all evaluated models.}
    \label{fig:heatmap-ctf-cybench}
\end{figure}

Despite exaggerated claims of full autonomy\footnote{There is a dangerous gap between automation and autonomy in cybersecurity~\cite{mayoral2025cybersecurity}. Organizations deploying mischaracterized \emph{autonomous} tools risk reducing oversight precisely when it is most needed, potentially creating new vulnerabilities.  As also described in \cite{kaliardos2022enough}, autonomy is typically treated as a system-level attribute, not a single component feature you \emph{bolt on}. In other words, we argue one does not magically get autonomy by swapping in an LLM; you get it when the overall system has delegated decision-making capability.} in cybersecurity, benchmarks are rapidly saturating~\cite{sanzgomez2025cybersecurityaibenchmarkcaibench} and as shown in Figures \ref{fig:heatmap-ctf-cybench} and \ref{fig:ctf-progression} (with a full model comparison in Figure~\ref{fig:heatmap-full}, Appendix~\ref{appendix:full-model-comparison}) popular cybersecurity benchmarks like Cybench~\cite{cyberbench2024} are rapidly being solved, with 50\%+ solved increase in the last 6 months in security-specialized LLMs, like the \texttt{alias} series of models\footnote{Refer to \cite{sanzgomez2025cybersecurityaibenchmarkcaibench} for comprehensive benchmarking methodology and evaluation results of the \texttt{alias} model series.}. The possibility of \emph{specialized} superintelligence is no longer speculative: as observed in \cite{y2025future}, ``we are running out of intelligence tests that humans can pass reliably and AI models cannot.'' If intelligence is essentially computational (the view held by most computational neuroscientists), then a working simulation of intelligence actually \emph{is} intelligence; it turned out to be a matter of scaling computation~\cite{y2025future}. This reframes the question from \emph{whether} machines can achieve superintelligent cybersecurity capabilities to \emph{how} to architect systems to reach that threshold.

%%%%%%%%%%%%%%%%%%%%%%%%%%%%%%%%%%%%%%%%%%%%%%%%%%%%%%%%%%%%%%%%%%%%%%%%%%%%%%%%%%%%%%%%

% Line plot showing CTF solve percentage over time by model series
\begin{figure}[!h]
    \centering
    \resizebox{\textwidth}{!}{%
    \begin{tikzpicture}
    % Define plot dimensions
    \def\plotwidth{14}
    \def\plotheight{6}
    \def\xmin{0}
    \def\xmax{26}  % Months from Feb 2024 to Apr 2026
    \def\ymin{0}
    \def\ymax{100}

    % Background
    \fill[white] (-1.5, -1) rectangle (\plotwidth+1, \plotheight+1);

    % Grid lines (horizontal)
    \foreach \y in {0, 20, 40, 60, 80, 100} {
        \pgfmathsetmacro{\ypos}{\y/100*\plotheight}
        \draw[graph_gray!30, thin] (0, \ypos) -- (\plotwidth, \ypos);
    }

    % Grid lines (vertical) - one per month label position
    \foreach \x in {0, 4, 8, 12, 16, 20, 24} {
        \pgfmathsetmacro{\xpos}{\x/26*\plotwidth}
        \draw[graph_gray!30, thin] (\xpos, 0) -- (\xpos, \plotheight);
    }

    % Axes
    \draw[graph_navy, thick] (0, 0) -- (\plotwidth, 0);
    \draw[graph_navy, thick] (0, 0) -- (0, \plotheight);

    % Y-axis labels
    \foreach \y in {0, 20, 40, 60, 80, 100} {
        \pgfmathsetmacro{\ypos}{\y/100*\plotheight}
        \node[font=\scriptsize\sffamily, text=graph_navy, anchor=east] at (-0.15, \ypos) {\y\%};
    }

    % Y-axis title
    \node[font=\small\sffamily, text=graph_navy, rotate=90, anchor=south] at (-1.1, \plotheight/2) {Cybench benchmark  (\% solved)};

    % X-axis labels (months)
    % Month 0 = Feb 2024, Month 24 = Feb 2026
    \node[font=\scriptsize\sffamily, text=graph_navy] at (0, -0.4) {Feb'24};
    \node[font=\scriptsize\sffamily, text=graph_navy] at ({4/26*\plotwidth}, -0.4) {Jun'24};
    \node[font=\scriptsize\sffamily, text=graph_navy] at ({8/26*\plotwidth}, -0.4) {Oct'24};
    \node[font=\scriptsize\sffamily, text=graph_navy] at ({12/26*\plotwidth}, -0.4) {Feb'25};
    \node[font=\scriptsize\sffamily, text=graph_navy] at ({16/26*\plotwidth}, -0.4) {Jun'25};
    \node[font=\scriptsize\sffamily, text=graph_navy] at ({20/26*\plotwidth}, -0.4) {Oct'25};
    \node[font=\scriptsize\sffamily, text=graph_navy] at ({24/26*\plotwidth}, -0.4) {Feb'26};

    % X-axis title
    \node[font=\small\sffamily, text=graph_navy] at (\plotwidth/2, -0.9) {Model Launch Date};

    % Convert date to x position: (month - Feb2024) / 24 * plotwidth
    % Feb 2024 = 0, Mar 2024 = 1, ..., Jan 2026 = 23

    % Brand-coherent colors (muted to highlight alias):
    % - Alias (Alias Robotics): cai_primary teal (HIGHLIGHTED)
    % - Claude (Anthropic): coral/terracotta - muted
    % - Gemini (Google): Google blue - muted
    % - GPT (OpenAI): grey - muted
    % - Mistral (Mistral AI): orange - muted
    \definecolor{anthropic_color}{RGB}{204, 119, 102}  % Anthropic coral/terracotta
    \definecolor{google_color}{RGB}{66, 133, 244}      % Google blue
    \definecolor{openai_color}{RGB}{100, 100, 100}     % Grey for OpenAI
    \definecolor{mistral_color}{RGB}{255, 140, 0}      % Mistral orange

    % === GEMINI PRO series (Google blue, muted) ===
    % gemini 1.5 pro: 02/24 (month 0), 6.1%
    % gemini 2.5 pro: 03/25 (month 13), 21.2%
    % gemini 3 pro: 11/25 (month 21), 63.6%
    \draw[google_color!50, thick, line width=1.5pt]
        ({0/26*\plotwidth}, {6.1/100*\plotheight}) --
        ({13/26*\plotwidth}, {21.2/100*\plotheight}) --
        ({21/26*\plotwidth}, {63.6/100*\plotheight});
    \fill[google_color!50] ({0/26*\plotwidth}, {6.1/100*\plotheight}) circle (3pt);
    \fill[google_color!50] ({13/26*\plotwidth}, {21.2/100*\plotheight}) circle (3pt);
    \fill[google_color!50] ({21/26*\plotwidth}, {63.6/100*\plotheight}) circle (3pt);
    % Labels
    \node[font=\tiny\sffamily, text=google_color!50, anchor=south west] at ({0/26*\plotwidth}, {6.1/100*\plotheight+0.1}) {gemini 1.5};
    \node[font=\tiny\sffamily, text=google_color!50, anchor=south west] at ({13/26*\plotwidth-0.5}, {21.2/100*\plotheight-0.5}) {gemini 2.5};
    \node[font=\tiny\sffamily, text=google_color!50, anchor=south east] at ({21/26*\plotwidth+0.4}, {63.6/100*\plotheight+0.1}) {gemini 3};

    % === CLAUDE OPUS series (Anthropic coral, muted) ===
    % claude 3 opus: 03/24 (month 1), 18.2%
    % claude opus 4.5: 11/25 (month 21), 81.8%
    % claude opus 4.6: 02/26 (month 24), 72.7%
    \draw[anthropic_color!70, thick, line width=1.5pt]
        ({1/26*\plotwidth}, {18.2/100*\plotheight}) --
        ({21/26*\plotwidth}, {81.8/100*\plotheight}) --
        ({24/26*\plotwidth}, {72.7/100*\plotheight});
    \fill[anthropic_color!70] ({1/26*\plotwidth}, {18.2/100*\plotheight}) circle (3pt);
    \fill[anthropic_color!70] ({21/26*\plotwidth}, {81.8/100*\plotheight}) circle (3pt);
    \fill[anthropic_color!70] ({24/26*\plotwidth}, {72.7/100*\plotheight}) circle (3pt);
    % Labels
    \node[font=\tiny\sffamily, text=anthropic_color!70, anchor=south west] at ({1/26*\plotwidth-0.1}, {18.2/100*\plotheight+0.15}) {opus 3};
    \node[font=\tiny\sffamily, text=anthropic_color!70, anchor=south] at ({21/26*\plotwidth}, {81.8/100*\plotheight+0.1}) {opus 4.5};
    \node[font=\tiny\sffamily, text=anthropic_color!70, anchor=south east] at ({24/26*\plotwidth+0.4}, {72.7/100*\plotheight-0.5}) {opus 4.6};

    % === CLAUDE SONNET series (Anthropic coral, muted) ===
    % claude 3.5 sonnet: 06/24 (month 4), 18.2%
    % claude 4 sonnet: 05/25 (month 15), 21.2%
    % claude sonnet 4.5: 09/25 (month 19), 48.5%
    \draw[anthropic_color!35, thick, line width=1.5pt]
        ({4/26*\plotwidth}, {18.2/100*\plotheight}) --
        ({15/26*\plotwidth}, {21.2/100*\plotheight}) --
        ({19/26*\plotwidth}, {48.5/100*\plotheight});
    \fill[anthropic_color!35] ({4/26*\plotwidth}, {18.2/100*\plotheight}) circle (3pt);
    \fill[anthropic_color!35] ({15/26*\plotwidth}, {21.2/100*\plotheight}) circle (3pt);
    \fill[anthropic_color!35] ({19/26*\plotwidth}, {48.5/100*\plotheight}) circle (3pt);
    % Labels
    \node[font=\tiny\sffamily, text=anthropic_color!35, anchor=north west] at ({4/26*\plotwidth-0.1}, {18.2/100*\plotheight+0.5}) {sonnet 3.5};
    \node[font=\tiny\sffamily, text=anthropic_color!35, anchor=south west] at ({15/26*\plotwidth-0.5}, {21.2/100*\plotheight+0.15}) {sonnet 4};
    \node[font=\tiny\sffamily, text=anthropic_color!35, anchor=south west] at ({19/26*\plotwidth+0.05}, {48.5/100*\plotheight+0.05}) {sonnet 4.5};

    % === GPT series (grey, muted) ===
    % gpt-4o: 05/24 (month 3), 15.2%
    % gpt 5: 08/25 (month 18), 33.3%
    % gpt 5.1: 11/25 (month 21), 39.4%
    % gpt 5.2: 12/25 (month 22), 48.5%
    \draw[openai_color!50, thick, line width=1.5pt]
        ({3/26*\plotwidth}, {15.2/100*\plotheight}) --
        ({18/26*\plotwidth}, {33.3/100*\plotheight}) --
        ({21/26*\plotwidth}, {39.4/100*\plotheight}) --
        ({22/26*\plotwidth}, {48.5/100*\plotheight});
    \fill[openai_color!50] ({3/26*\plotwidth}, {15.2/100*\plotheight}) circle (3pt);
    \fill[openai_color!50] ({18/26*\plotwidth}, {33.3/100*\plotheight}) circle (3pt);
    \fill[openai_color!50] ({21/26*\plotwidth}, {39.4/100*\plotheight}) circle (3pt);
    \fill[openai_color!50] ({22/26*\plotwidth}, {48.5/100*\plotheight}) circle (3pt);
    % Labels
    \node[font=\tiny\sffamily, text=openai_color!50, anchor=north] at ({3/26*\plotwidth-0.45}, {15.2/100*\plotheight+0.3}) {gpt 4o};
    \node[font=\tiny\sffamily, text=openai_color!50, anchor=south west] at ({18/26*\plotwidth+0}, {33.3/100*\plotheight-0.45}) {gpt 5};
    \node[font=\tiny\sffamily, text=openai_color!50, anchor=north] at ({21/26*\plotwidth}, {39.4/100*\plotheight-0.2}) {gpt 5.1};
    \node[font=\tiny\sffamily, text=openai_color!50, anchor=south west] at ({22/26*\plotwidth+0.1}, {48.5/100*\plotheight+0.05}) {gpt 5.2};

    % === MISTRAL LARGE series (Mistral orange, muted) ===
    % mistral large 2.0: 07/24 (month 5), 12.1% (4 solved)
    % mistral large 2.1: 11/24 (month 9), 3.0% (1 solved)
    % mistral large 3: 12/25 (month 22), 12.1% (4 solved)
    \draw[mistral_color!50, thick, line width=1.5pt]
        ({5/26*\plotwidth}, {12.1/100*\plotheight}) --
        ({9/26*\plotwidth}, {3.0/100*\plotheight}) --
        ({22/26*\plotwidth}, {12.1/100*\plotheight});
    \fill[mistral_color!50] ({5/26*\plotwidth}, {12.1/100*\plotheight}) circle (3pt);
    \fill[mistral_color!50] ({9/26*\plotwidth}, {3.0/100*\plotheight}) circle (3pt);
    \fill[mistral_color!50] ({22/26*\plotwidth}, {12.1/100*\plotheight}) circle (3pt);
    % Labels
    \node[font=\tiny\sffamily, text=mistral_color!50, anchor=south] at ({5/26*\plotwidth}, {12.1/100*\plotheight-0.6}) {large 2.0};
    \node[font=\tiny\sffamily, text=mistral_color!50, anchor=north] at ({9/26*\plotwidth}, {3.0/100*\plotheight+0.6}) {large 2.1};
    \node[font=\tiny\sffamily, text=mistral_color!50, anchor=north west] at ({22/26*\plotwidth-0.3}, {12.1/100*\plotheight-0.2}) {large 3};

    % === ALIAS series (CAI teal - Alias Robotics) ===
    % alias0: 05/25 (month 15), 18.2%
    % alias1: 10/25 (month 20), 42.4%
    % alias2: 01/26 (month 23), 75.8%
    \draw[cai_primary, thick, line width=1.5pt]
        ({15/26*\plotwidth}, {18.2/100*\plotheight}) --
        ({20/26*\plotwidth}, {42.4/100*\plotheight}) --
        ({23/26*\plotwidth}, {75.8/100*\plotheight});
    \fill[cai_primary] ({15/26*\plotwidth}, {18.2/100*\plotheight}) circle (3pt);
    \fill[cai_primary] ({20/26*\plotwidth}, {42.4/100*\plotheight}) circle (3pt);
    \fill[cai_primary] ({23/26*\plotwidth}, {75.8/100*\plotheight}) circle (3pt);
    % Labels
    \node[font=\tiny\sffamily, text=cai_primary, anchor=north] at ({15/26*\plotwidth}, {18.2/100*\plotheight-0.15}) {alias0};
    \node[font=\tiny\sffamily, text=cai_primary, anchor=south west] at ({20/26*\plotwidth+0.1}, {42.4/100*\plotheight-0.1}) {alias1};
    \node[font=\tiny\sffamily, text=cai_primary, anchor=south west] at ({23/26*\plotwidth-0.3}, {75.8/100*\plotheight+0.15}) {alias2};

    % Legend with company names
    \node[font=\scriptsize\sffamily\bfseries, text=graph_navy, anchor=west] at (\plotwidth+0.3, \plotheight-0.2) {Model Series};

    % Alias entry
    \draw[cai_primary, thick, line width=1.5pt] (\plotwidth+0.3, \plotheight-0.7) -- (\plotwidth+0.8, \plotheight-0.7);
    \fill[cai_primary] (\plotwidth+0.55, \plotheight-0.7) circle (3pt);
    \node[font=\scriptsize\sffamily, text=graph_navy, anchor=west] at (\plotwidth+0.9, \plotheight-0.7) {Alias Robotics' alias};

    % Other entries (muted)
    \draw[anthropic_color!70, thick, line width=1.5pt] (\plotwidth+0.3, \plotheight-1.2) -- (\plotwidth+0.8, \plotheight-1.2);
    \fill[anthropic_color!70] (\plotwidth+0.55, \plotheight-1.2) circle (3pt);
    \node[font=\scriptsize\sffamily, text=graph_navy!60, anchor=west] at (\plotwidth+0.9, \plotheight-1.2) {Anthropic's claude opus};

    \draw[anthropic_color!35, thick, line width=1.5pt] (\plotwidth+0.3, \plotheight-1.7) -- (\plotwidth+0.8, \plotheight-1.7);
    \fill[anthropic_color!35] (\plotwidth+0.55, \plotheight-1.7) circle (3pt);
    \node[font=\scriptsize\sffamily, text=graph_navy!60, anchor=west] at (\plotwidth+0.9, \plotheight-1.7) {Anthropic's claude sonnet};

    \draw[google_color!50, thick, line width=1.5pt] (\plotwidth+0.3, \plotheight-2.2) -- (\plotwidth+0.8, \plotheight-2.2);
    \fill[google_color!50] (\plotwidth+0.55, \plotheight-2.2) circle (3pt);
    \node[font=\scriptsize\sffamily, text=graph_navy!60, anchor=west] at (\plotwidth+0.9, \plotheight-2.2) {Google's gemini pro};

    \draw[openai_color!50, thick, line width=1.5pt] (\plotwidth+0.3, \plotheight-2.7) -- (\plotwidth+0.8, \plotheight-2.7);
    \fill[openai_color!50] (\plotwidth+0.55, \plotheight-2.7) circle (3pt);
    \node[font=\scriptsize\sffamily, text=graph_navy!60, anchor=west] at (\plotwidth+0.9, \plotheight-2.7) {OpenAI's gpt};

    \draw[mistral_color!50, thick, line width=1.5pt] (\plotwidth+0.3, \plotheight-3.2) -- (\plotwidth+0.8, \plotheight-3.2);
    \fill[mistral_color!50] (\plotwidth+0.55, \plotheight-3.2) circle (3pt);
    \node[font=\scriptsize\sffamily, text=graph_navy!60, anchor=west] at (\plotwidth+0.9, \plotheight-3.2) {Mistral AI's large};

    \end{tikzpicture}%
    }% end resizebox
    \caption{\texttt{Cybench} solve rate progression over time by model series, highlighting the \texttt{alias} series. The x-axis shows model launch dates, y-axis shows solved percentage of \texttt{CAIBench-Jeopardy CTFs (Cybench)}~\cite{sanzgomez2025cybersecurityaibenchmarkcaibench} benchmark. Each experiment was run for a maximum of 300 agentic interactions, 245 minutes of computing time per challenge, a maximum of 40 USD per challenge on API model expenses and with \emph{pass@3}. Plot depicts how most models are rapidly improving, showing signs of benchmark saturation. A comprehensive comparison of all evaluated models is provided in Figure \ref{fig:heatmap-full} at Appendix~\ref{appendix:full-model-comparison}.}
    \label{fig:ctf-progression}
\end{figure}

%%%%%%%%%%%%%%%%%%%%%%%%%%%%%%%%%%%%%%%%%%%%%%%%%%%%%%%%

Bostrom~\cite{bostrom2014superintelligence} defines superintelligence as ``an intellect that greatly exceeds the cognitive performance of humans in virtually all domains of interest.'' We argue that \emph{Cybersecurity Superintelligence} merits treatment as a distinct domain-specific instantiation. While no standardized term exists, superhuman cyber capability is increasingly discussed in technical and policy research: Bengio et al.~\cite{bengio2025international} identify cybersecurity as a key ``dangerous capability'' domain; Hendrycks et al.~\cite{hendrycks2025superintelligence} analyze national security implications of AI systems that ``can be turned to destructive ends''; and Potter et al.~\cite{potter2025frontier} marginal-risk modeling and position analyses argue frontier AI exerts stronger influence on cyber offense than defense, though Balassone et al.~\cite{balassone2025cybersecurity} empirically refute this, demonstrating no statistical offensive advantage when defenders also use AI.

The domain-specific treatment is justified by cybersecurity's unique structural complexity. Cybersecurity demands simultaneous mastery across heterogeneous disciplines (reverse engineering, binary exploitation, cryptanalysis, forensics, hardware security), each with distinct toolchains and reasoning modalities. For a superintelligence, these must be applied across fundamentally different contexts: IT, OT/ICS, robotics, IoT, mobile, each introducing domain-specific protocols and threat models. This combinatorial complexity (disciplines $\times$ sectors $\times$ constraints) creates an evaluation space no human can fully cover. The most ``superhuman-adjacent'' capability to date: LLM-driven agents operating 3,600$\times$ faster than humans in specific security tasks (11$\times$ overall), while winning worldwide competitions with \$50,000+ in prizes during 2025~\cite{aliasrobotics2025cai, mayoral2025worldstopagent}.

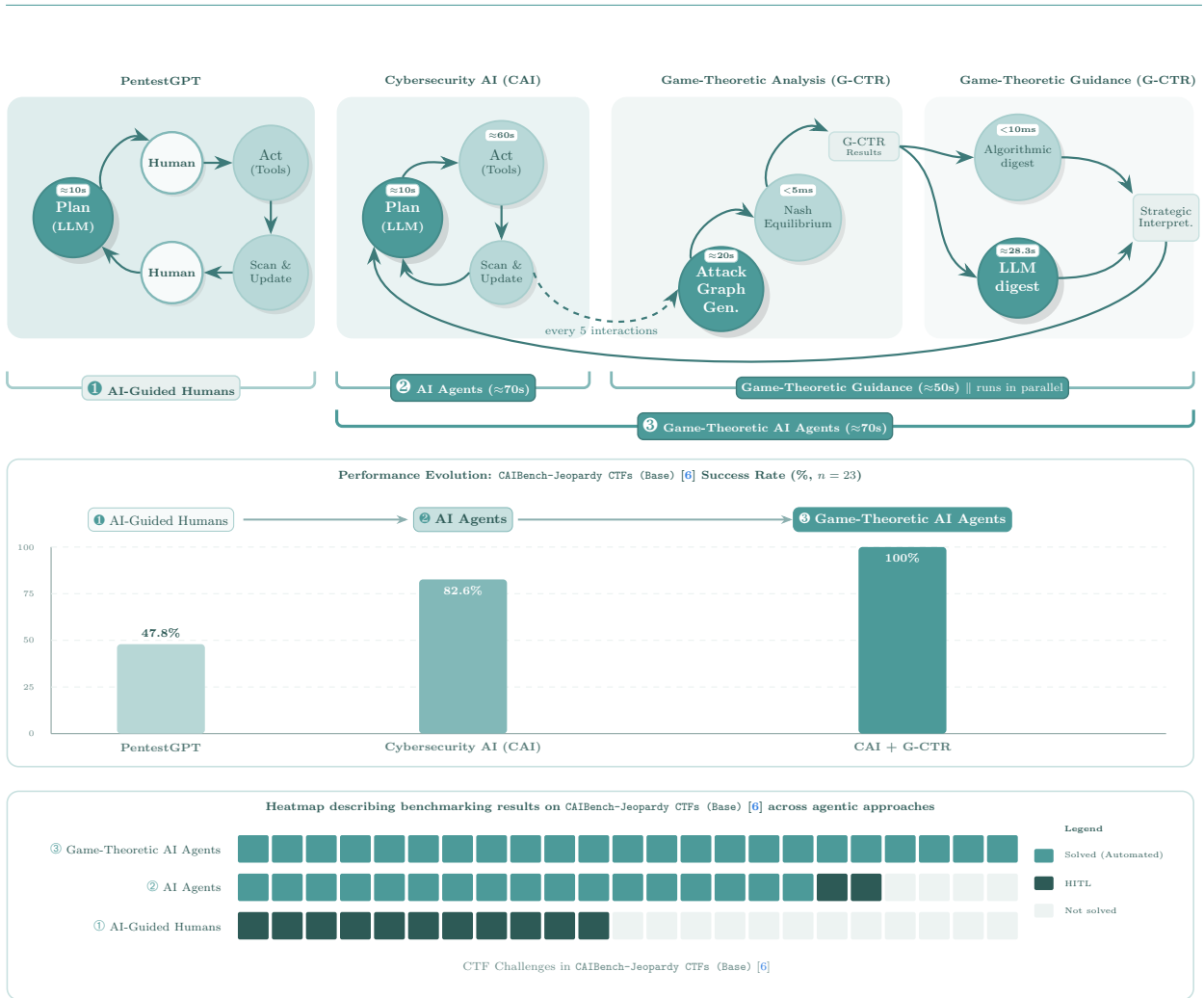
\begin{figure}[h!]
    \centering
    \resizebox{\textwidth}{!}{%
    \begin{tikzpicture}[scale=1.0, every node/.style={scale=0.85, font=\sffamily},
        datanode/.style={circle, draw=graph_teal!50, fill=graph_lightcyan, minimum size=1.3cm, align=center, font=\scriptsize, text=graph_navy, line width=0.9pt, drop shadow={opacity=0.18, shadow xshift=0.08cm, shadow yshift=-0.08cm}},
        keynode/.style={circle, draw=graph_teal!90!black, fill=graph_teal, text=white, minimum size=1.7cm, align=center, font=\small\bfseries, line width=1pt, drop shadow={opacity=0.25, shadow xshift=0.12cm, shadow yshift=-0.12cm}},
        humannode/.style={circle, draw=graph_teal!70, fill=graph_human, minimum size=0.9cm, align=center, font=\tiny\bfseries, text=graph_navy, line width=1.2pt, drop shadow={opacity=0.15, shadow xshift=0.06cm, shadow yshift=-0.06cm}},
        supportnode/.style={rectangle, draw=graph_lightcyan!80!black, fill=graph_lightcyan, minimum width=1.6cm, minimum height=0.45cm, align=center, font=\tiny, rounded corners=2pt, line width=0.6pt},
        altnode/.style={rectangle, draw=graph_teal!30, fill=graph_gray, minimum width=1.5cm, minimum height=0.45cm, align=center, font=\tiny, text=graph_navy, rounded corners=3pt, line width=0.7pt},
    ]

    \fill[graph_teal!18, rounded corners=8pt] (-13.6, -2.2) rectangle (-8.0, 2.2);
    \node[font=\scriptsize\bfseries, text=graph_navy] at (-10.8, 2.5) {\textsc{PentestGPT}};

    % Agent Execution (LEFT) - medium saturation
    \fill[graph_teal!12, rounded corners=8pt] (-7.6, -2.2) rectangle (-3.0, 2.2);
    \node[font=\scriptsize\bfseries, text=graph_navy] at (-5.3, 2.5) {\textsc{Cybersecurity AI} (CAI)};

    % Game-Theoretic Analysis (MIDDLE) - lighter
    \fill[graph_gray!60, rounded corners=8pt] (-2.6, -2.2) rectangle (2.7, 2.2);
    \node[font=\scriptsize\bfseries, text=graph_navy] at (0.4, 2.5) {\textsc{Game-Theoretic Analysis} (G-CTR)};

    % Digest Generation (RIGHT) - lightest
    \fill[graph_gray!40, rounded corners=8pt] (3.1, -2.2) rectangle (8.0, 2.2);
    \node[font=\scriptsize\bfseries, text=graph_navy] at (5.9, 2.5) {\textsc{Game-Theoretic Guidance} (G-CTR)};

    % ============================================================================
    % LLM-Assisted Execution (FAR LEFT) - with human intervention points
    % ============================================================================
    \node[keynode] (hagent) at (-12.4, 0) {\small Plan\\ \scriptsize(LLM)};
    \node[humannode] (human1) at (-10.6, 1.0) {\scriptsize Human};
    \node[datanode, minimum width=1.6cm] (hactions) at (-8.8, 1.0) {\small Act\\ \scriptsize(Tools)};
    \node[datanode, minimum width=1.6cm] (hobserve) at (-8.8, -1.0) {\scriptsize Scan \&\\Update};
    \node[humannode] (human2) at (-10.6, -1.0) {\scriptsize Human};

    % LLM-Assisted arrows - cleaner paths avoiding node overlaps
    \draw[-{Stealth[scale=1.2]}, line width=1.1pt, graph_arrow]
        (hagent.north east) to[out=45, in=180, bend left=65] (human1.north west);
    \draw[-{Stealth[scale=1.2]}, line width=1.1pt, graph_arrow]
        (human1.east) -- (hactions.west);
    \draw[-{Stealth[scale=1.2]}, line width=1.1pt, graph_arrow]
        (hactions.south) -- (hobserve.north);
    \draw[-{Stealth[scale=1.2]}, line width=1.1pt, graph_arrow]
        (hobserve.west) -- (human2.east);
    \draw[-{Stealth[scale=1.2]}, line width=1.1pt, graph_arrow]
        (human2.west) to[out=180, in=-45] (hagent.south east);

    % ============================================================================
    % Agent Execution (LEFT)
    % ============================================================================
    \node[keynode] (agent) at (-6.4, 0) {\small Plan\\ \scriptsize(LLM)};
    \node[datanode, minimum width=1.8cm] (actions) at (-4.6, 1.0) {\small Act\\ \scriptsize(Tools)};
    \node[datanode] (observe) at (-4.6, -1.0) {\scriptsize Scan \&\\Update};

    % Agent Execution arrows
    \draw[-{Stealth[scale=1.2]}, line width=1.1pt, graph_arrow]
        (agent.north) to[out=60, in=180] (actions.west);
    \draw[-{Stealth[scale=1.2]}, line width=1.1pt, graph_arrow]
        (actions.south) to[out=-90, in=90] (observe.north);
    \draw[-{Stealth[scale=1.2]}, line width=1.1pt, graph_arrow]
        (observe.west) to[out=180, in=-60, bend left=65] (agent.south);

    % ============================================================================
    % Game-Theoretic Analysis (MIDDLE)
    % ============================================================================
    \node[keynode] (graph) at (-0.6, -1.3) {\small Attack\\Graph\\Gen.};
    \node[datanode] (nash) at (0.8, 0) {\scriptsize Nash\\Equilibrium};
    \node[altnode] (data) at (2.0, 1.3) {\scriptsize G-CTR\\Results};

    % G-CTR Analysis arrows
    \draw[-{Stealth[scale=1.2]}, line width=1.1pt, graph_arrow]
        (graph.north west) to[out=60, in=-120, bend left=65] (nash.west);
    \draw[-{Stealth[scale=1.2]}, line width=1.1pt, graph_arrow]
        (nash.north west) to[out=60, in=-120, bend left=65] (data.north west);

    % ============================================================================
    % Digest Generation (RIGHT)
    % ============================================================================
    \node[datanode] (algo) at (4.8, 1.1) {\scriptsize Algorithmic\\ digest};
    \node[keynode] (llm) at (4.8, -1.1) {\small LLM \\digest};
    \node[altnode, minimum size=1.0cm] (digest) at (7.5, 0) {\scriptsize Strategic\\\scriptsize Interpret.};

    % From G-CTR to Digest Generation
    \draw[-{Stealth[scale=1.2]}, line width=1.1pt, graph_arrow]
        (data.east) to[out=0, in=180] (algo.west);
    \draw[-{Stealth[scale=1.2]}, line width=1.1pt, graph_arrow]
        (data.east) to[out=-30, in=150] (llm.west);

    % Digest Generation convergence
    \draw[-{Stealth[scale=1.2]}, line width=1.1pt, graph_arrow]
        (algo.east) to[out=0, in=120] (digest.north west);
    \draw[-{Stealth[scale=1.2]}, line width=1.1pt, graph_arrow]
        (llm.east) to[out=0, in=-120] (digest.south west);

    % ============================================================================
    % Cross-section arrows
    % ============================================================================
    % From Agent Execution to G-CTR Analysis
    \draw[-{Stealth[scale=1.2]}, line width=1.1pt, dashed, graph_arrow]
        (observe.east) to[out=0, in=180, bend right=65] node[pos=0.5, below, font=\scriptsize, inner sep=4pt] {every 5 interactions} (graph.west);

    % Feedback loop from Digest back to Agent
    \draw[-{Stealth[scale=1.2]}, line width=1.1pt, graph_arrow]
        (digest.south) to[out=-90, in=-90, looseness=0.5] (agent.south west);

    % ============================================================================
    % Timing annotations - cohesive styling
    % ============================================================================
    % HITL timing
    \node[font=\tiny\bfseries, text=graph_navy, fill=white, inner sep=2pt, rounded corners=2pt, draw=graph_teal!40, line width=0.4pt]
        at (-12.4, 0.5) {$\approx$10s};

    % Agent Execution timing
    \node[font=\tiny\bfseries, text=graph_navy, fill=white, inner sep=2pt, rounded corners=2pt, draw=graph_teal!40, line width=0.4pt]
        at (-6.4, 0.5) {$\approx$10s};
    \node[font=\tiny\bfseries, text=graph_navy, fill=white, inner sep=2pt, rounded corners=2pt, draw=graph_teal!40, line width=0.4pt]
        at (-4.6, 1.5) {$\approx$60s};

    % G-CTR Analysis timing
    \node[font=\tiny\bfseries, text=graph_navy, fill=white, inner sep=2pt, rounded corners=2pt, draw=graph_teal!40, line width=0.4pt]
        at (-0.6, -0.7) {$\approx$20s};
    \node[font=\tiny\bfseries, text=graph_navy, fill=white, inner sep=2pt, rounded corners=2pt, draw=graph_teal!40, line width=0.4pt]
        at (0.8, 0.5) {$<$5ms};

    % Digest Generation timing
    \node[font=\tiny\bfseries, text=graph_navy, fill=white, inner sep=2pt, rounded corners=2pt, draw=graph_teal!40, line width=0.4pt]
        at (4.8, 1.6) {$<$10ms};
    \node[font=\tiny\bfseries, text=graph_navy, fill=white, inner sep=2pt, rounded corners=2pt, draw=graph_teal!40, line width=0.4pt]
        at (4.8, -0.6) {$\approx$28.3s};

    % % Parallel execution indicator between Phase 3 and Phase 1
    % \draw[line width=0.8pt, graph_navy!40] (-3.0, -1.8) -- (-3.0, 1.8);
    % \draw[line width=0.8pt, graph_navy!40] (-2.9, -1.8) -- (-2.9, 1.8);
  
    % % Bidirectional arrows showing parallel flow
    % \draw[-{Stealth[scale=0.8]}, line width=0.8pt, graph_teal] (3.45, 1.4) -- (3.45, 0.8);
    % \draw[-{Stealth[scale=0.8]}, line width=0.8pt, graph_teal] (3.45, -0.8) -- (3.45, -1.4);
  
    % Parallel symbol and label
    % \node[font=\tiny\bfseries, text=white, fill=graph_teal, inner sep=2pt, rounded corners=2pt, draw=graph_teal!80!black, line width=0.5pt]
    %     at (3.45, 0) {$\parallel$};
    % \node[font=\tiny, text=graph_navy!80, rotate=90]
    %     at (3.45, -0.5) {async};
    % \node[font=\tiny, text=graph_navy!80, rotate=90]
    %     at (3.45, 0.5) {parallel};
  
    % Time budget indicators at bottom - LLM-Assisted (FAR LEFT)
    \draw[line width=1.4pt, graph_teal!50, rounded corners=3pt] (-13.6, -2.8) -- (-13.6, -3.1) -- (-8.0, -3.1) -- (-8.0, -2.8);
    \node[font=\scriptsize\bfseries, text=graph_navy, fill=graph_gray, inner sep=3pt, rounded corners=3pt, draw=graph_teal!40, line width=0.7pt]
        at (-10.8, -3.1) {\textcolor{cai_primary}{{\large\ding{182}}} AI-Guided Humans};

    % Time budget indicators at bottom - Agent Execution (LEFT): ~70s
    \draw[line width=1.4pt, graph_teal!70, rounded corners=3pt] (-7.6, -2.8) -- (-7.6, -3.1) -- (-3.0, -3.1) -- (-3.0, -2.8);
    \node[font=\scriptsize\bfseries, text=white, fill=graph_teal, inner sep=3pt, rounded corners=3pt, draw=graph_teal!80!black, line width=0.7pt]
        at (-5.3, -3.1) {{\large\ding{183}} AI Agents ($\approx$70s)};

    % Time budget indicators at bottom - G-CTR + Digest (RIGHT): ~50s
    \draw[line width=1.4pt, graph_teal!70, rounded corners=3pt] (-2.6, -2.8) -- (-2.6, -3.1) -- (8.0, -3.1) -- (8.0, -2.8);
    \node[font=\scriptsize\bfseries, text=white, fill=graph_teal, inner sep=3pt, rounded corners=3pt, draw=graph_teal!80!black, line width=0.7pt]
        at (2.7, -3.1) {Game-Theoretic Guidance ($\approx$50s) {\normalfont\scriptsize\textcolor{white!70}{$\parallel$ runs in parallel}}};

    % ============================================================================
    % Grouping bracket for AI Agents + Game-Theoretic Guidance (both teal sections)
    % ============================================================================
    \draw[line width=1.6pt, graph_teal, rounded corners=3pt] (-7.6, -3.5) -- (-7.6, -3.8) -- (8.0, -3.8) -- (8.0, -3.5);
    \node[font=\scriptsize\bfseries, text=white, fill=graph_teal, inner sep=3pt, rounded corners=3pt, draw=graph_teal!80!black, line width=0.7pt]
        at (0.2, -3.8) {{\large\ding{184}} Game-Theoretic AI Agents ($\approx$70s)};

    % ============================================================================
    % Performance Evolution Chart - Below the architecture (shifted down for grouping bracket)
    % ============================================================================

    % Chart background (increased vertical gap from grouping bracket)
    \fill[white, rounded corners=6pt] (-13.6, -9.8) rectangle (8.0, -4.4);
    \draw[graph_teal!30, line width=0.8pt, rounded corners=6pt] (-13.6, -10.0) rectangle (8.0, -4.4);

    % Chart title
    \node[font=\scriptsize\bfseries, text=graph_navy] at (-2.8, -4.7) {\textsc{Performance Evolution: \texttt{CAIBench-Jeopardy CTFs (Base)}~\cite{sanzgomez2025cybersecurityaibenchmarkcaibench} Success Rate (\%, $n=23$)}};

    % Y-axis (range from -9.4 to -6.0, giving 3.4 units for 100%)
    \draw[graph_navy!50, line width=0.6pt] (-12.8, -9.4) -- (-12.8, -6.0);
    % Y-axis labels
    \node[font=\tiny, text=graph_navy!70, anchor=east] at (-13.0, -9.4) {0};
    \node[font=\tiny, text=graph_navy!70, anchor=east] at (-13.0, -8.55) {25};
    \node[font=\tiny, text=graph_navy!70, anchor=east] at (-13.0, -7.7) {50};
    \node[font=\tiny, text=graph_navy!70, anchor=east] at (-13.0, -6.85) {75};
    \node[font=\tiny, text=graph_navy!70, anchor=east] at (-13.0, -6.0) {100};
    % Y-axis grid lines
    \draw[graph_gray, line width=0.3pt, dashed] (-12.8, -8.55) -- (7.5, -8.55);
    \draw[graph_gray, line width=0.3pt, dashed] (-12.8, -7.7) -- (7.5, -7.7);
    \draw[graph_gray, line width=0.3pt, dashed] (-12.8, -6.85) -- (7.5, -6.85);
    \draw[graph_gray, line width=0.3pt, dashed] (-12.8, -6.0) -- (7.5, -6.0);

    % X-axis
    \draw[graph_navy!50, line width=0.6pt] (-12.8, -9.4) -- (7.5, -9.4);

    % === BAR CHART - Three groups aligned with architecture sections ===

    % Bar 1: LLM-Assisted (PentestGPT) - 11/23 = 47.8% success rate
    \fill[graph_teal!40, rounded corners=2pt] (-11.6, -9.4) rectangle (-10.0, -7.77);
    \node[font=\scriptsize\bfseries, text=graph_navy] at (-10.8, -7.57) {47.8\%};
    \node[font=\scriptsize\bfseries, text=graph_navy!80, align=center] at (-10.8, -9.65) {PentestGPT};

    % Bar 2: Agent Execution (CAI) - 19/23 = 82.6% success rate
    \fill[graph_teal!70, rounded corners=2pt] (-6.1, -9.4) rectangle (-4.5, -6.59);
    \node[font=\scriptsize\bfseries, text=white] at (-5.3, -6.79) {82.6\%};
    \node[font=\scriptsize\bfseries, text=graph_navy!80, align=center] at (-5.3, -9.65) {Cybersecurity AI (CAI)};

    % Bar 3: G-CTR + Digest (CAI + G-CTR) - 23/23 = 100% success rate
    \fill[graph_teal, rounded corners=2pt] (1.9, -9.4) rectangle (3.5, -6.0);
    \node[font=\scriptsize\bfseries, text=white] at (2.7, -6.2) {100\%};
    \node[font=\scriptsize\bfseries, text=graph_navy!80, align=center] at (2.7, -9.65) {CAI + G-CTR};

    % % === TREND LINE - Showing progression ===
    % \draw[graph_arrow, line width=1.8pt, -{Stealth[scale=1.0]}]
    %     (-10.8, -7.21) to[out=30, in=150] (-5.3, -6.59) to[out=30, in=150] (2.7, -6.03);

    % === IMPROVEMENT ANNOTATIONS ===
    % % Arrow and label: PentestGPT to CAI
    % \draw[graph_teal!60, line width=0.8pt, -{Stealth[scale=0.8]}] (-8.8, -6.9) -- (-6.8, -6.65);
    % \node[font=\tiny, text=graph_teal!80!black, fill=white, inner sep=1.5pt, rounded corners=1pt] at (-7.8, -6.55) {+64\%};

    % % Arrow and label: CAI to CAI-GT
    % \draw[graph_teal!60, line width=0.8pt, -{Stealth[scale=0.8]}] (-2.3, -6.35) -- (0.4, -6.15);
    % \node[font=\tiny, text=graph_teal!80!black, fill=white, inner sep=1.5pt, rounded corners=1pt] at (-0.9, -6.05) {+34\%};

    % === ICONS/LABELS for each stage ===
    % Human icon area (LLM-Assisted)
    \node[rectangle, rounded corners=3pt, draw=graph_teal!50, fill=graph_human, minimum height=0.5cm, font=\tiny, text=graph_navy] at (-10.8, -5.5) { {\small\cding{182}} \footnotesize AI-Guided Humans};

    % AI icon area (Agent Execution)
    \node[rectangle, rounded corners=3pt, draw=graph_teal!70, fill=graph_teal!30, minimum size=0.5cm, font=\tiny\bfseries, text=graph_navy] at (-5.3, -5.5) {{\small\cding{183}} \footnotesize AI Agents};

    % AI+ icon area (G-CTR guided)
    \node[rectangle, rounded corners=3pt, draw=graph_teal, fill=graph_teal, minimum size=0.5cm, font=\tiny\bfseries, text=white] at (2.7, -5.5) {{\small\ding{184}} \footnotesize Game-Theoretic AI Agents};

    % Connecting evolution arrows between icons
    \draw[graph_arrow!60, line width=1pt, -{Stealth[scale=0.8]}] (-9.3, -5.5) -- (-6.3, -5.5);
    \draw[graph_arrow!60, line width=1pt, -{Stealth[scale=0.8]}] (-4.3, -5.5) -- (0.7, -5.5);

    % ============================================================================
    % HEATMAP - CTF Scenario Comparison across three approaches
    % ============================================================================

    % Heatmap background (wider to accommodate y-axis labels and legend)
    \fill[white, rounded corners=6pt] (-13.6, -14.25) rectangle (8.0, -10.45);
    \draw[graph_teal!30, line width=0.8pt, rounded corners=6pt] (-13.6, -14.25) rectangle (8.0, -10.45);

    % Heatmap title
    \node[font=\scriptsize\bfseries, text=graph_navy] at (-2.8, -10.75) {\textsc{Heatmap describing benchmarking results on \texttt{CAIBench-Jeopardy CTFs (Base)}~\cite{sanzgomez2025cybersecurityaibenchmarkcaibench} across agentic approaches}};

    % Y-axis labels (three approaches) - inside the boundary
    \node[font=\scriptsize, text=graph_navy, anchor=east] at (-9.6, -11.5) {{\small\cding{194}} Game-Theoretic AI Agents};
    \node[font=\scriptsize, text=graph_navy, anchor=east] at (-9.6, -12.2) {{\small\cding{193}} AI Agents};
    \node[font=\scriptsize, text=graph_navy, anchor=east] at (-9.6, -12.9) {{\small\cding{192}} AI-Guided Humans};

    % Heatmap grid - 23 columns x 3 rows
    % Cell colors: graph_teal (solved), graph_gray (not solved), graph_navy (HITL)

    % Define cell dimensions - 23 cells from x=-9.4 to x=5.0
    \pgfmathsetmacro{\cellw}{0.62}
    \pgfmathsetmacro{\cellh}{0.55}
    \pgfmathsetmacro{\startx}{-9.4}

    % Row 1: Game-Theoretic AI Agents (highest success - mostly solved) - 23 elements
    \foreach \i/\status in {0/1,1/1,2/1,3/1,4/1,5/1,6/1,7/1,8/1,9/1,10/1,11/1,12/1,13/1,14/1,15/1,16/1,17/1,18/1,19/1,20/1,21/1,22/1} {
        \pgfmathsetmacro{\xpos}{\startx + \i*\cellw}
        \ifnum\status=1
            \fill[graph_teal, rounded corners=1pt] (\xpos, -11.75) rectangle (\xpos+\cellw-0.06, -11.25);
        \else
            \fill[graph_gray!80, rounded corners=1pt] (\xpos, -11.75) rectangle (\xpos+\cellw-0.06, -11.25);
        \fi
    }

    % Row 2: AI Agents (medium success) - 23 elements
    \foreach \i/\status in {0/1,1/1,2/1,3/1,4/1,5/1,6/1,7/1,8/1,9/1,10/1,11/1,12/1,13/1,14/1,15/1,16/1,17/2,18/2,19/0,20/0,21/0,22/0} {
        \pgfmathsetmacro{\xpos}{\startx + \i*\cellw}
        \ifnum\status=1
            \fill[graph_teal, rounded corners=1pt] (\xpos, -12.45) rectangle (\xpos+\cellw-0.06, -11.95);
        \else
            \ifnum\status=2
                \fill[graph_navy, rounded corners=1pt] (\xpos, -12.45) rectangle (\xpos+\cellw-0.06, -11.95);
            \else
                \fill[graph_gray!80, rounded corners=1pt] (\xpos, -12.45) rectangle (\xpos+\cellw-0.06, -11.95);
            \fi
        \fi
    }

    % Row 3: AI-Guided Humans (lowest autonomous success, some HITL) - 23 elements
    \foreach \i/\status in {0/2,1/2,2/2,3/2,4/2,5/2,6/2,7/2,8/2,9/2,10/2,11/0,12/0,13/0,14/0,15/0,16/0,17/0,18/0,19/0,20/0,21/0,22/0} {
        \pgfmathsetmacro{\xpos}{\startx + \i*\cellw}
        \ifnum\status=1
            \fill[graph_teal, rounded corners=1pt] (\xpos, -13.15) rectangle (\xpos+\cellw-0.06, -12.65);
        \else
            \ifnum\status=2
                \fill[graph_navy, rounded corners=1pt] (\xpos, -13.15) rectangle (\xpos+\cellw-0.06, -12.65);
            \else
                \fill[graph_gray!80, rounded corners=1pt] (\xpos, -13.15) rectangle (\xpos+\cellw-0.06, -12.65);
            \fi
        \fi
    }

    % X-axis label
    \node[font=\scriptsize, text=graph_navy!70] at (-2.5, -13.65) {CTF Challenges in \texttt{CAIBench-Jeopardy CTFs (Base)}~\cite{sanzgomez2025cybersecurityaibenchmarkcaibench}};

    % Legend - positioned to the left, inside boundary
    \node[font=\tiny\bfseries, text=graph_navy] at (6.0, -11.15) {Legend};
    \fill[graph_teal, rounded corners=1pt] (5.1, -11.5) rectangle (5.45, -11.75);
    \node[font=\tiny, text=graph_navy, anchor=west] at (5.55, -11.625) {Solved (Automated)};

    \fill[graph_navy, rounded corners=1pt] (5.1, -12.0) rectangle (5.45, -12.25);
    \node[font=\tiny, text=graph_navy, anchor=west] at (5.55, -12.125) {HITL};

    \fill[graph_gray!80, rounded corners=1pt] (5.1, -12.5) rectangle (5.45, -12.75);
    \node[font=\tiny, text=graph_navy, anchor=west] at (5.55, -12.625) {Not solved};

    \end{tikzpicture}%
    }% end resizebox

    \caption{\small{Progression towards Cybersecurity Superintelligence: From AI-Guided Humans to Game-Theoretic AI Agents.} The architecture illustrates three evolutionary stages: \cding{182}~\textbf{AI-Guided Humans} (PentestGPT, far left): LLMs provide planning assistance while humans remain in the loop for action execution and observation interpretation, achieving 47.8\% success rate. \cding{183} Human expert-level \textbf{AI Agents} (CAI, center-left): Cybersecurity AI agents automating the security testing process and leading to 82.6\% success rate. \cding{184}~\textbf{Game-Theoretic AI Agents} (CAI + G-CTR, right): game-theoretic reasoning augments the agent via attack graph generation, Nash equilibrium computation, and strategic digest injection, achieving 100\% success rate on the same benchmark. The bar chart (middle) quantifies performance gains across stages, while the heatmap (bottom) shows per-challenge resolution, demonstrating that game-theoretic guidance enables solving challenges that pure AI agents cannot.}
    \label{fig:system_architecture}
  \end{figure}

In this work, we present an evolution toward Cybersecurity Superintelligence. We trace this evolution through three landmark achievements (Figure~\ref{fig:system_architecture}) that mark a paradigmatic shift: we begin by \cding{182} using AI to augment human capabilities (Section \ref{sub:pentestgpt}), followed by \cding{183} using AI to build human expert-level agents (Section \ref{sub:cai}), and culminate in \cding{184} using human reasoning to guide AI toward superhuman performance (Section \ref{sub:gctr}). This progression—from AI-guided humans to human-guided AI—represents a fundamental shift in the role each (human and AI) plays, where human expertise transitions from actor to operator, and from operator to supervisor. We examine this inversion and its implications in detail.

\section{Evolution Toward Superintelligence}\label{sec:evolution}

%%%%%%%%%%%%%%%%%%%%%%%%%%%%%%%%%%%%%%%%%%%%%%%%%%%%%%%%
% G-CTR Architecture Figure (from CAI-GT paper)
%%%%%%%%%%%%%%%%%%%%%%%%%%%%%%%%%%%%%%%%%%%%%%%%%%%%%%%%

\subsection{AI-Guided Humans: PentestGPT}\label{sub:pentestgpt}

PentestGPT~\cite{deng2024pentestgpt} pioneered LLM-assisted penetration testing through the first systematic evaluation of LLM capabilities in offensive security, with concurrent independent work by Happe et al.~\cite{happe2023getting} similarly exploring LLM-driven penetration testing approaches. Benchmarking across 182 sub-tasks, PentestGPT revealed that while LLMs excel at discrete operations (tool configuration, output interpretation, vulnerability identification), they fail at coherent multi-step strategies due to context loss from token constraints, recency bias toward immediate tasks, and hallucination-induced inaccuracies.

PentestGPT's architecture (Figure~\ref{fig:system_architecture}, \cding{182}) addresses these limitations through module separation inspired by penetration testing team dynamics. The internal \emph{Reasoning Module} maintains global context via the Penetration Testing Task Tree (PTT), an attributed tree $T = (N, A)$ encoding testing status in natural language, with verification preventing hallucination-induced structural corruption. The \emph{Generation Module} translates sub-tasks into executable commands via Chain-of-Thought decomposition, isolating tactical execution from strategic context. The \emph{Parsing Module} condenses verbose tool outputs into actionable advice.

Humans remain central as command executors, output validators, and strategic correctors. This inverts traditional expertise requirements: the LLM encodes domain knowledge (vulnerability patterns, exploitation techniques, tool configurations) while humans provide tool execution and judgment. Effectively, PentestGPT empowered \emph{AI-guided humans} to conduct penetration testing by augmenting their capabilities with expert-level reasoning, democratizing offensive security and enabling users with limited background to leverage sophisticated intuition through natural language guidance.

Achieving 228.6\% improvement over baseline GPT-3.5 and placing 24th among 248 CTF teams, PentestGPT (6,500+ GitHub stars; adopted by AWS, Huawei, TikTok) validated human-AI collaboration. However, its reliance on human tool execution revealed the bottleneck CAI would address.

%%%%%%%%%%%%%%%%%%%%%%%%%%%%%%%%%%%%%%%%%%%%%%%%%%%%%%%%%%%%%%%%%%%%%%
\subsection{Expert-Level Agents: CAI}\label{sub:cai}

\begin{table}[!h]
    \centering
    \small
    \renewcommand{\arraystretch}{1.3}
    \setlength{\tabcolsep}{3pt}
    \begin{tabular*}{\textwidth}{@{\extracolsep{\fill}}lrrrrrr@{}}
        \toprule
        \textbf{Category} & \textcolor{cai_color}{\textbf{$\sum t_{\text{CAI}}$} (s)} & \textcolor{cai_color}{\textbf{$\sum{c_{\text{CAI}}}$} (\$)} & \textcolor{human_color}{\textbf{$\sum{t_{\text{Human}}}$} (s)} & \textcolor{human_color}{\textbf{$\sum{c_{\text{Human}}}$} (\$)} & \textbf{$t_{ratio}$} & \textcolor{cai_color}{\textbf{\timecolor{$c_{ratio}$}}} \\
        \midrule
        rev        & \textcolor{cai_color}{\textbf{541}} (9m 1s)       & \textcolor{cai_color}{\textbf{0.83}} & \textcolor{human_color}{418789} (4d 20h)  & \textcolor{human_color}{5642} & 774x & \timecolor{6797x} \\
        misc       & \textcolor{cai_color}{\textbf{1650}} (27m 30s)              & \textcolor{cai_color}{\textbf{3.04}} & \textcolor{human_color}{38364} (10h 39m)       & \textcolor{human_color}{516} & 23x & \timecolor{169x} \\
        pwn        & \textcolor{cai_color}{99368} (1d 3h)              & \textcolor{cai_color}{93} & \textcolor{human_color}{\textbf{77407}} (21h 30m) & \textcolor{human_color}{\textbf{1042}} & 0.77x & \timecolor{11x} \\
        web        & \textcolor{cai_color}{\textbf{558}} (9m 18s)     & \textcolor{cai_color}{\textbf{1.78}} & \textcolor{human_color}{31264} (8h 41m)    & \textcolor{human_color}{421} & 56x & \timecolor{236x} \\
        crypto     & \textcolor{cai_color}{9549} (2h 39m)            & \textcolor{cai_color}{2.03} & \textcolor{human_color}{\textbf{4483}} (1h 14m) & \textcolor{human_color}{\textbf{60}} & 0.47x & \timecolor{29x} \\
        forensics  & \textcolor{cai_color}{\textbf{432}} (7m 12s)      & \textcolor{cai_color}{\textbf{1.78}} & \textcolor{human_color}{405361} (4d 16h)  & \textcolor{human_color}{5461} & 938x & \timecolor{3067x} \\
        robotics   & \textcolor{cai_color}{\textbf{408}} (6m 48s)     & \textcolor{cai_color}{\textbf{6.6}} & \textcolor{human_color}{302400} (3d 12h)     & \textcolor{human_color}{4074} & 741x & \timecolor{617x} \\
        \midrule
        \textbf{$\sum$} & \textcolor{cai_color}{\textbf{112506}} (1d 7h)  & \textcolor{cai_color}{\textbf{109}} & \textcolor{human_color}{1278068} (14d 19h) & \textcolor{human_color}{17218} & 11x  & \timecolor{156x}\\
        \bottomrule
    \end{tabular*}
    \caitablecaption{Comparison of the sum of time ($t$), cost ($c$) and respective ratios of CAI and Human performance across different CTF challenge categories. Each row shows the sum of average completion times and costs for all challenges within that category, for both CAI and Human participants. CAI cost corresponds with the API expenses. Human cost was calculated using the hourly rates of \euro{45} (\$48.54)~\cite{aliasrobotics2025cai}. For the sake of readability, for \textbf{$t_{ratio}$} and \textbf{$c_{ratio}$}, values under 10 were rounded to two decimals (rounding up the third decimal). Values $\geq{10}$ were rounded to the nearest integer.  Best performance (lower time/cost) per category is \textbf{bolded}. Values in parentheses represent human-readable time formats. The bottom row shows the total sum across all categories, representing the cumulative performance difference. The summary given here is fully expanded in detail in \cite{aliasrobotics2025cai}.}
    \label{tab:category_table}
\end{table}

\begin{table}[!h]
    \centering
    \small
    \renewcommand{\arraystretch}{1.3}
    \setlength{\tabcolsep}{3pt}
    \begin{tabular*}{\textwidth}{@{\extracolsep{\fill}}lrrrrrr@{}}
        \toprule
        \textbf{Difficulty} & \textcolor{cai_color}{\textbf{$\sum$}\textbf{${t_{\text{CAI}}}$} (s)} & \textcolor{cai_color}{\textbf{$\sum$}\textbf{${c_{\text{CAI}}}$} (\$)} & \textcolor{human_color}{\textbf{$\sum$}\textbf{${t_{\text{Human}}}$} (s)} & \textcolor{human_color}{\textbf{$\sum$}\textbf{${c_{\text{Human}}}$} (\$)} & \textbf{$t_{ratio}$} & \textcolor{cai_color}{\textbf{\timecolor{$c_{ratio}$}}} \\
        \midrule
        Very Easy & \textcolor{cai_color}{\textbf{1067}} (17m 46s) & \textcolor{cai_color}{\textbf{3.02}} & \textcolor{human_color}{852765} (9d 20h) & \textcolor{human_color}{11488} & 799x & \timecolor{3803x} \\
        Easy & \textcolor{cai_color}{26463} (7h 21m) & \textcolor{cai_color}{43} & \textcolor{human_color}{\textbf{25879}} (7h 11m) & \textcolor{human_color}{\textbf{348}} & 0.98x & \timecolor{8.03x} \\
        Medium & \textcolor{cai_color}{\textbf{29821}} (8h 16m) & \textcolor{cai_color}{\textbf{41}} & \textcolor{human_color}{353704} (4d 2h) & \textcolor{human_color}{4765} & 11x & \timecolor{115x} \\
        Hard & \textcolor{cai_color}{37935} (10h 32m) & \textcolor{cai_color}{6.88} & \textcolor{human_color}{\textbf{34569}} (9h 36m) & \textcolor{human_color}{\textbf{465}} & 0.91x & \timecolor{68x} \\
        Insane & \textcolor{cai_color}{17220} (4h 47m) & \textcolor{cai_color}{15} & \textcolor{human_color}{\textbf{11151}} (3h 5m) & \textcolor{human_color}{\textbf{150}} & 0.65x & \timecolor{9.79x} \\
        \bottomrule
    \end{tabular*}
    \caitablecaption{Comparison of the sum of time ($t$), cost ($c$) and respective ratios of CAI and Human performance across difficulty levels.}
    \label{tab:difficulty_table}
\end{table}

CAI~\cite{aliasrobotics2025cai} eliminated PentestGPT's human tool execution bottleneck through a fully automated agent-centric architecture framework. Where PentestGPT required humans to execute commands and validate outputs,\footnote{As of January 2026, PentestGPT v1.0.0 has evolved into an agentic tool capable of conducting automated tasks without human intervention, though CAI pioneered this automated approach.} CAI allows building expert-level agents that operate end-to-end: reasoning via LLMs, executing through integrated tools, and adapting based on results, all without human intervention.

The CAI framework (Figure~\ref{fig:system_architecture}, \cding{183}) comprises six architectural pillars: \texttt{Agents} (specialized security actors), \texttt{Tools} (command execution, web interaction, code manipulation), \texttt{Handoffs} (inter-agent control transfer), \texttt{Patterns} (collaborative agent topologies like Swarm for red team operations), \texttt{Turns} (interaction cycle management), and \texttt{HITL} (optional human oversight). This modular design enables specialized agents (red team, bug bounty hunter, blue team) to coordinate through well-defined handoff protocols, dynamically shifting expertise as new information emerges.

Benchmarking across 54 CTF challenges against human experts revealed dramatic performance asymmetries (Table~\ref{tab:category_table}). CAI achieved 774× speedup in reverse engineering (9 minutes vs.\ 4.8 days), 938× in forensics (7 minutes vs.\ 4.7 days), and 741× in robotics challenges, domains requiring pattern recognition and systematic enumeration where AI parallelism excels. Conversely, humans outperformed CAI in pwn (0.77×) and crypto (0.47×), categories demanding creative exploitation and mathematical insight that current LLMs handle less effectively. Difficulty-level analysis (Table~\ref{tab:difficulty_table}) shows CAI dominating ``Very Easy'' challenges (799× faster) while approaching parity at higher difficulties, suggesting LLM limitations in long-horizon planning and novel attack synthesis.

The cost differential proved equally stark: \$109 total API cost versus \$17,218 equivalent human labor (156× reduction). Beyond benchmarks, CAI demonstrated competitive dominance across the 2025 CTF circuit \cite{mayoral2025worldstopagent}: Rank \#6 at Dragos OT CTF (1,200+ teams), \#1 at Neurogrid CTF claiming the \$50,000 prize (41/45 flags), \#1 among AI teams in HTB ``AI vs Human'' (\$750 award), \#22 peak at Cyber Apocalypse (8,129 teams), and \#21 at UWSP Pointer Overflow (635 teams), consistently solving challenges 37\% faster than elite human teams. Yet this dominance exposed a fundamental limitation: CAI matched or exceeded human \emph{speed}, but not human \emph{strategic reasoning}. The transition from expert-level to superintelligent performance requires agents that reason about adversarial dynamics, the game-theoretic intuition that distinguishes elite security professionals from the average hacker.

%%%%%%%%%%%%%%%%%%%%%%%%%%%%%%%%%%%%%%%%%%%%%%%%%%%%%%%%%%%%%%%%%%%%%%
\subsection{Game-Theoretic Agents: G-CTR}\label{sub:gctr}

\begin{figure}[h!]
    \centering
    \resizebox{\textwidth}{!}{%
    \begin{tikzpicture}[font=\sffamily]
    
        % ============================================================================
        % LEFT: Attack Graph Image (60%)
        % ============================================================================
    
        % Graph container
        \fill[white, rounded corners=6pt] (-8.5, -4.0) rectangle (2.5, 5.0);
        \draw[graph_teal!40, line width=0.8pt, rounded corners=6pt] (-8.5, -4.0) rectangle (2.5, 5.0);
    
        % Title
        \node[font=\scriptsize\bfseries, text=graph_navy, fill=white, inner sep=2pt, rounded corners=2pt] at (-3.0, 4.6) {Attack Graph Topology};
    
        % Image placeholder - actual image will be included
        \node[anchor=center] at (-3.0, 0.5) {\includegraphics[width=10.5cm, height=7.5cm, keepaspectratio]{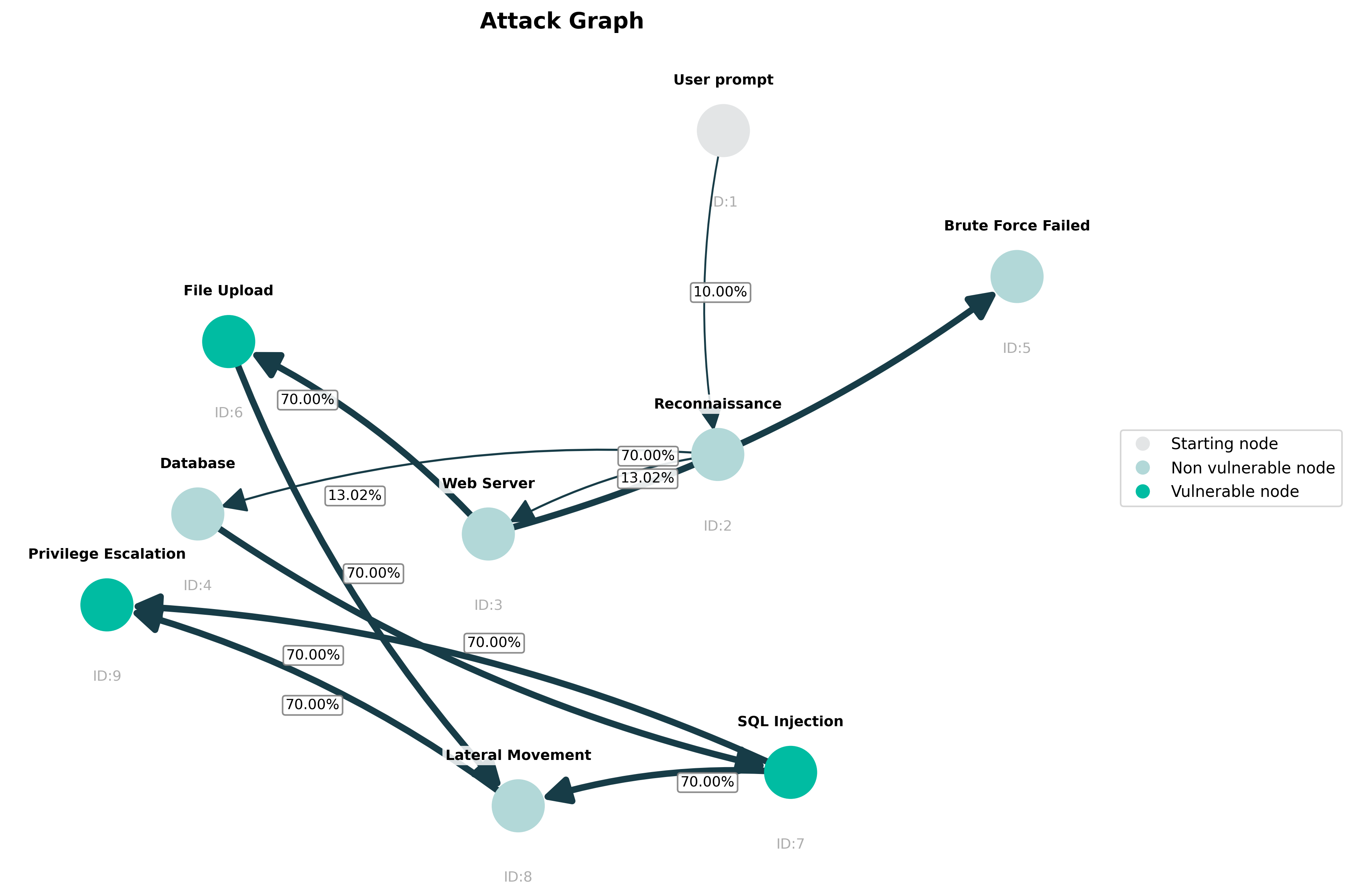}};
    
        % ============================================================================
        % RIGHT: Strategy Tables (40%)
        % ============================================================================
    
        % Right panel container
        \fill[white, rounded corners=6pt] (3.0, -4.0) rectangle (10.5, 5.0);
        \draw[graph_teal!40, line width=0.8pt, rounded corners=6pt] (3.0, -4.0) rectangle (10.5, 5.0);
    
        % === Defense Strategy Table (Top) ===
        \node[font=\scriptsize\bfseries, text=graph_navy] at (6.75, 4.6) {Nash Equilibrium Strategies};
    
        % Defense table background
        \fill[graph_teal!15, rounded corners=4pt] (3.3, 2.2) rectangle (10.2, 4.3);
    
        % Defense table header
        \fill[graph_teal, rounded corners=3pt] (3.4, 3.8) rectangle (10.1, 4.2);
        \node[font=\fontsize{8}{9}\selectfont\bfseries, text=white] at (6.75, 4.0) {Defense Strategy (Node Allocation)};

        % Defense table content
        \node[font=\fontsize{7}{8}\selectfont\bfseries, text=graph_navy] at (4.5, 3.4) {Node};
        \node[font=\fontsize{7}{8}\selectfont\bfseries, text=graph_navy] at (6.0, 3.4) {Prob.};
        \node[font=\fontsize{7}{8}\selectfont\bfseries, text=graph_navy] at (7.8, 3.4) {Node};
        \node[font=\fontsize{7}{8}\selectfont\bfseries, text=graph_navy] at (9.3, 3.4) {Prob.};

        \draw[graph_teal!40, line width=0.4pt] (3.5, 3.15) -- (10.0, 3.15);

        % Data rows
        \node[font=\fontsize{7}{8}\selectfont, text=graph_navy] at (4.5, 2.85) {8};
        \node[font=\fontsize{7}{8}\selectfont\bfseries, text=graph_teal] at (6.0, 2.85) {0.674};
        \node[font=\fontsize{7}{8}\selectfont, text=graph_navy] at (7.8, 2.85) {2};
        \node[font=\fontsize{7}{8}\selectfont, text=graph_navy!60] at (9.3, 2.85) {0.000};

        \node[font=\fontsize{7}{8}\selectfont, text=graph_navy] at (4.5, 2.45) {4};
        \node[font=\fontsize{7}{8}\selectfont\bfseries, text=graph_teal] at (6.0, 2.45) {0.326};
        \node[font=\fontsize{7}{8}\selectfont, text=graph_navy] at (7.8, 2.45) {3};
        \node[font=\fontsize{7}{8}\selectfont, text=graph_navy!60] at (9.3, 2.45) {0.000};
    
        % === Attacker Strategy Table (Middle) ===
        \fill[graph_teal!10, rounded corners=4pt] (3.3, -1.35) rectangle (10.2, 2.0);

        % Attacker table header
        \fill[graph_teal!80, rounded corners=3pt] (3.4, 1.5) rectangle (10.1, 1.9);
        \node[font=\fontsize{8}{9}\selectfont\bfseries, text=white] at (6.75, 1.7) {Attack Strategy (Path Selection)};

        % Attacker table headers
        \node[font=\fontsize{6.5}{7.5}\selectfont\bfseries, text=graph_navy] at (3.8, 1.15) {ID};
        \node[font=\fontsize{6.5}{7.5}\selectfont\bfseries, text=graph_navy, anchor=west] at (4.3, 1.15) {Path};
        \node[font=\fontsize{6.5}{7.5}\selectfont\bfseries, text=graph_navy] at (9.6, 1.15) {Prob.};

        \draw[graph_teal!40, line width=0.4pt] (3.5, 0.9) -- (10.0, 0.9);

        % Attack paths
        \node[font=\fontsize{6.5}{7.5}\selectfont, text=graph_navy] at (3.8, 0.55) {5};
        \node[font=\fontsize{6.5}{7.5}\selectfont, text=graph_navy, anchor=west] at (4.3, 0.55) {1$\rightarrow$2$\rightarrow$4$\rightarrow$7$\rightarrow$9};
        \node[font=\fontsize{6.5}{7.5}\selectfont\bfseries, text=graph_teal] at (9.6, 0.55) {0.674};

        \node[font=\fontsize{6.5}{7.5}\selectfont, text=graph_navy] at (3.8, 0.15) {1};
        \node[font=\fontsize{6.5}{7.5}\selectfont, text=graph_navy, anchor=west] at (4.3, 0.15) {1$\rightarrow$2$\rightarrow$3$\rightarrow$6$\rightarrow$8$\rightarrow$9};
        \node[font=\fontsize{6.5}{7.5}\selectfont\bfseries, text=graph_teal] at (9.6, 0.15) {0.326};

        \node[font=\fontsize{6.5}{7.5}\selectfont, text=graph_navy!60] at (3.8, -0.25) {3};
        \node[font=\fontsize{6.5}{7.5}\selectfont, text=graph_navy!60, anchor=west] at (4.3, -0.25) {1$\rightarrow$2$\rightarrow$4$\rightarrow$7};
        \node[font=\fontsize{6.5}{7.5}\selectfont, text=graph_navy!60] at (9.6, -0.25) {0.000};

        \node[font=\fontsize{6.5}{7.5}\selectfont, text=graph_navy!60] at (3.8, -0.65) {2};
        \node[font=\fontsize{6.5}{7.5}\selectfont, text=graph_navy!60, anchor=west] at (4.3, -0.65) {1$\rightarrow$2$\rightarrow$4};
        \node[font=\fontsize{6.5}{7.5}\selectfont, text=graph_navy!60] at (9.6, -0.65) {0.000};

        \node[font=\fontsize{6.5}{7.5}\selectfont, text=graph_navy!60] at (3.8, -1.0) {4};
        \node[font=\fontsize{6.5}{7.5}\selectfont, text=graph_navy!60, anchor=west] at (4.3, -1.0) {1$\rightarrow$2$\rightarrow$4$\rightarrow$7$\rightarrow$8$\rightarrow$9};
        \node[font=\fontsize{6.5}{7.5}\selectfont, text=graph_navy!60] at (9.6, -1.0) {0.000};
    
        % === Game Equilibrium Box (Bottom) ===
        \fill[graph_teal, rounded corners=5pt] (3.3, -3.85) rectangle (10.2, -1.7);

        % Equilibrium title
        \node[font=\fontsize{9}{10}\selectfont\bfseries, text=white] at (6.75, -2.0) {Game Equilibrium (Nash)};

        \draw[white, line width=0.4pt] (3.6, -2.3) -- (9.9, -2.3);

        % Defender metric
        \node[font=\fontsize{7}{8}\selectfont, text=white!90, anchor=west] at (3.6, -2.55) {Defender success threshold:};
        \node[font=\fontsize{9}{10}\selectfont\bfseries, text=white, anchor=east] at (9.9, -2.55) {3.528\%};
        \node[font=\fontsize{5}{6}\selectfont\itshape, text=white!70, anchor=west] at (3.8, -2.8) {Defender can keep attacker success below this value};

        % Attacker metric
        \node[font=\fontsize{7}{8}\selectfont, text=white!90, anchor=west] at (3.6, -3.2) {Attacker guaranteed success:};
        \node[font=\fontsize{9}{10}\selectfont\bfseries, text=white, anchor=east] at (9.9, -3.2) {3.528\%};
        \node[font=\fontsize{5}{6}\selectfont\itshape, text=white!70, anchor=west] at (3.8, -3.45) {Attacker can guarantee at least this success probability};
    
        % % Equilibrium indicator
        % \node[font=\fontsize{5}{6}\selectfont\itshape, text=white!80] at (6.75, -3.5) {Saddle point $\Rightarrow$ Optimal mixed strategies};
    
    \end{tikzpicture}%
    }
    \caption{\small{Game-Theoretic Attack Graph Analysis.} Left: Attack graph topology showing nodes (vulnerabilities) and edges (attack transitions) extracted from the LLM context. Right: Nash equilibrium strategies computed by G-CTR algorithm. A good defense strategy would allocate monitoring resources to nodes 8 (67.4\%) and 4 (32.6\%), while optimal attack paths would exploit through nodes 1$\rightarrow$2$\rightarrow$4$\rightarrow$7$\rightarrow$9 and 1$\rightarrow$2$\rightarrow$3$\rightarrow$6$\rightarrow$8$\rightarrow$9, yielding an equilibrium success probability of 3.528\%. Refer to \cite{mayoralvilches2025gametheoretic} for more details.}
    \label{fig:Graph_attack_example}
\end{figure}
    %https://drive.google.com/drive/folders/1dleDyJZKncvkfJHonPHhCAM47EeD1eT9?usp=drive_link

CAI's expert-level performance revealed a fundamental ceiling: speed and autonomy alone do not constitute superintelligence in cybersecurity. Matching human experts, even at 3,600× their speed, still produces human-equivalent reasoning. Surpassing human capability requires agents that reason strategically, the way humans mentally \emph{play the game}. Just as a chess grandmaster evaluates attacker/defender lines before committing to a move, security professionals apply game theory: evaluating the current state, imagining adversary responses, and choosing actions that maximize long-term advantage. This strategic reasoning, not faster execution, separates expert-level from superhuman performance.

G-CTR \cite{mayoralvilches2025gametheoretic} addresses this gap through a neurosymbolic architecture that embeds game-theoretic reasoning into LLM-based agents' system prompt. Rather than relying solely on pattern-matched intuitions prone to hallucination and logical inconsistency, agents consult explicit payoff computations and equilibrium analyses, a symbolic scaffold grounding actions in principled adversarial reasoning. As Jones et al.~\cite{jones2025nature} observe, ``using rules-based systems during crucial reasoning steps can help keep LLMs from going off-track''; G-CTR instantiates this principle for cybersecurity through three phases (Figure~\ref{fig:system_architecture}, \cding{184}).

The architecture operates via closed-loop strategic feedback. First, \textbf{Attack Graph Generation} extracts structured graph representations from unstructured security logs (or raw LLM context) using LLMs, achieving 70--90\% node correspondence with expert annotations\footnote{Validation was performed by two professional security researchers with no affiliation to this research, hired as independent evaluators.} while running 60--245× faster. Second, \textbf{Nash Equilibrium Computation} applies the Cut-the-Rope (CTR)~\cite{ctr_paper1,ctr_paper2,rass2025poster} algorithm to identify optimal attack/defense strategies; Figure~\ref{fig:Graph_attack_example} illustrates how G-CTR computes defense allocations (nodes 8: 67.4\%, node 4: 32.6\%) and attack path probabilities yielding a 3.528\% equilibrium success rate. Third, \textbf{Strategic Digest Injection} transforms equilibrium computations into natural language guidance inserted into the agent's system prompt, steering subsequent actions toward statistically advantageous continuations. This digest (akin to a chess engine highlighting strongest lines) reduces ambiguity, collapses the search space, and suppresses hallucinations by anchoring the model to what is actually unfolding.

% \begin{figure}[ht]
% \centering
% \includegraphics[width=0.9\columnwidth]{core/2.cai-gt/img/attack_graph_simplified.png}
% \caption{\textbf{Game-theoretic attack graph extraction.} G-CTR computes Nash equilibria for optimal attack/defense strategies.}
% \label{fig:attack-graph}
% \end{figure}

Empirical validation~\cite{mayoralvilches2025gametheoretic} across 44 cyber-range penetration tests demonstrates the strategic advantage: success rates doubled (20.0\%→42.9\%), cost-per-success decreased (and thereby improved) 2.7× (\$0.32→\$0.12), and behavioral variance reduced 5.2×, indicating more consistent, predictable agent behavior. In Attack \& Defense scenarios, the best configuration (experimentally determined) was to let red and blue agents, which mirror the actions of real-life red teams that attack a system and blue teams that defend it, share a common attack graph as their joint battle-field. This configuration defeated LLM-only baselines 2:1 and outperformed independently-guided dual teams 3.7:1. These results demonstrate that game-theoretic guidance transforms expert-level agents into strategically superior ones: not merely faster AI agents, but systems exhibiting reasoning capabilities that exceed humans while maintaining mathematical rigor in adversarial decision-making. We argue that these game-theoretic agents pave the way towards cybersecurity superintelligence.

\section{Discussion}\label{sec:discussion}
%This work documents the emergence of cybersecurity superintelligence through three pivotal contributions that have pioneered the field of AI Security. PentestGPT \cite{deng2024pentestgpt} established that LLMs can guide humans through complex penetration testing workflows, achieving 228.6\% improvement over baseline models. CAI \cite{aliasrobotics2025cai} demonstrated that automated agents can match expert-level human performance while operating 3,600× faster and reducing costs 156-fold, validated through \#1 rankings at Neurogrid CTF (\$50,000 prize), Dragos OT, and HTB ``AI vs Human'' competitions \cite{mayoral2025worldstopagent}. G-CTR \cite{mayoralvilches2025gametheoretic} introduced neurosymbolic reasoning that doubles success rates, reduces behavioral variance 5.2×, and achieves 2:1 advantage over non-strategic AI. Together, these advances establish a clear trajectory: from AI-guided humans (\cding{182}) to expert-level automated agents (\cding{183}) to game-theoretic superintelligence (\cding{184}), as illustrated in Figure~\ref{fig:system_architecture}.

%%%%%%%%%%%%%%%%%%%%%%%%%%%%%%%%%%%%%%%%%%%%%%%%%%%%%%%%
% Cybersecurity Superintelligence Figure - Clean Table Design
%%%%%%%%%%%%%%%%%%%%%%%%%%%%%%%%%%%%%%%%%%%%%%%%%%%%%%%%
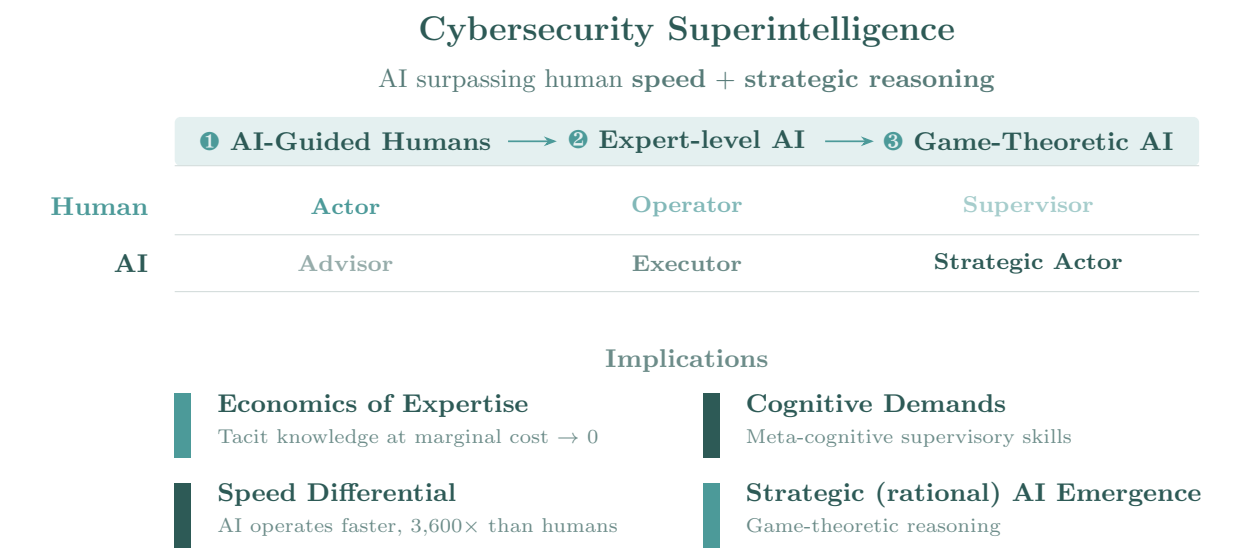
\begin{figure}[b!]
    \centering
    \resizebox{0.95\textwidth}{!}{%
    \begin{tikzpicture}[every node/.style={font=\sffamily}]

    % ============================================================================
    % TOP: Title and Definition
    % ============================================================================
    \node[font=\large\bfseries, text=graph_navy] at (0, 3.6) {Cybersecurity Superintelligence};
    \node[font=\small, text=graph_navy!80] at (0, 3.0) {AI surpassing human \textbf{speed} + \textbf{strategic reasoning}};

    % ============================================================================
    % MIDDLE: Role Inversion Table
    % ============================================================================

    % Column header background
    \fill[graph_teal!15, rounded corners=2pt] (-6.3, 2.55) rectangle (6.3, 1.95);

    % Column headers
    \node[font=\small\bfseries, text=graph_navy] at (-4.2, 2.25) {\cding{182} AI-Guided Humans};
    \node[font=\small\bfseries, text=graph_navy] at (0, 2.25) {\cding{183} Expert-level AI};
    \node[font=\small\bfseries, text=graph_navy] at (4.2, 2.25) {\cding{184} Game-Theoretic AI};

    % Progression arrows
    \draw[-{Stealth[scale=0.8]}, line width=0.6pt, graph_teal] (-2.2, 2.25) -- (-1.6, 2.25);
    \draw[-{Stealth[scale=0.8]}, line width=0.6pt, graph_teal] (1.7, 2.25) -- (2.3, 2.25);

    % Row labels
    \node[font=\small\bfseries, text=graph_teal, anchor=east] at (-6.5, 1.45) {Human};
    \node[font=\small\bfseries, text=graph_navy, anchor=east] at (-6.5, 0.75) {AI};

    % Table lines
    \draw[graph_navy!20, line width=0.4pt] (-6.3, 1.95) -- (6.3, 1.95);
    \draw[graph_navy!20, line width=0.4pt] (-6.3, 1.1) -- (6.3, 1.1);
    \draw[graph_navy!20, line width=0.4pt] (-6.3, 0.4) -- (6.3, 0.4);

    % Human row - emphasize the transition with decreasing weight
    \node[font=\footnotesize\bfseries, text=graph_teal] at (-4.2, 1.45) {{Actor}};
    \node[font=\footnotesize\bfseries, text=graph_teal!70] at (0, 1.45) {{Operator}};
    \node[font=\footnotesize\bfseries, text=graph_teal!50] at (4.2, 1.45) {{Supervisor}};

    % AI row - emphasize the transition with increasing weight
    \node[font=\footnotesize\bfseries, text=graph_navy!50] at (-4.2, 0.75) {Advisor};
    \node[font=\footnotesize\bfseries, text=graph_navy!70] at (0, 0.75) {Executor};
    \node[font=\footnotesize\bfseries, text=graph_navy] at (4.2, 0.75) {Strategic Actor};

    % ============================================================================
    % BOTTOM: Four Implications - Clean Layout
    % ============================================================================

    % \draw[graph_teal!40, line width=0.5pt] (-6.3, -0.1) -- (6.3, -0.1);
    \node[font=\small\bfseries, text=graph_navy!70] at (0, -0.45) {\textsc{Implications}};

    % 2x2 grid - accent bars only, no dots

    % Row 1
    \fill[graph_teal] (-6.3, -0.85) rectangle (-6.1, -1.65);
    \node[font=\small\bfseries, text=graph_navy, anchor=west] at (-5.9, -1.0) {Economics of Expertise};
    \node[font=\scriptsize, text=graph_navy!70, anchor=west] at (-5.9, -1.4) {Tacit knowledge at marginal cost $\rightarrow$ 0};

    \fill[graph_navy] (0.2, -0.85) rectangle (0.4, -1.65);
    \node[font=\small\bfseries, text=graph_navy, anchor=west] at (0.6, -1.0) {Cognitive Demands};
    \node[font=\scriptsize, text=graph_navy!70, anchor=west] at (0.6, -1.4) {Meta-cognitive supervisory skills};

    % Row 2
    \fill[graph_navy] (-6.3, -1.95) rectangle (-6.1, -2.75);
    \node[font=\small\bfseries, text=graph_navy, anchor=west] at (-5.9, -2.1) {Speed Differential};
    \node[font=\scriptsize, text=graph_navy!70, anchor=west] at (-5.9, -2.5) {AI operates faster, 3,600$\times$ than humans};

    \fill[graph_teal] (0.2, -1.95) rectangle (0.4, -2.75);
    \node[font=\small\bfseries, text=graph_navy, anchor=west] at (0.6, -2.1) {Strategic (rational) AI Emergence};
    \node[font=\scriptsize, text=graph_navy!70, anchor=west] at (0.6, -2.5) {Game-theoretic reasoning};

    \end{tikzpicture}
    }
    \caption{\small{Role inversion in cybersecurity superintelligence.} Humans transition from Actor (executing tasks with AI advice) to Supervisor (overseeing AI strategy). Conversely, AI evolves from Advisor to Strategic Actor, assuming both execution and game-theoretic reasoning. This inversion redefines expertise economics, cognitive demands, operational tempo, and strategic dynamics.}
    \label{fig:role_inversion}
\end{figure}

The progression, from AI-guided humans (Section \ref{sub:pentestgpt}) to game-theoretic AI agents to human-guided AI (Section \ref{sub:gctr}), represents a fundamental inversion in the relationship between human expertise and machine capability (Figure~\ref{fig:role_inversion}). %In the initial paradigm (\cding{182}), AI served as a cognitive prosthesis: LLMs externalized domain knowledge into natural language guidance while humans retained executive control over action selection, tool invocation, and strategic decision-making. The human remained the \emph{actor}, the primary agent executing security operations, while AI functioned as an enhanced reference system. With automated AI agents (\cding{183}), this relationship inverts: AI takes on the actor role, executing end-to-end penetration testing workflows while humans transition to \emph{operators}, prompting objectives, configuring parameters, defining scope, and monitoring execution without direct involvement in tactical decisions. The emergence of game-theoretic reasoning (\cding{184}) completes this transformation: as AI systems demonstrate strategic reasoning that anticipates adversarial responses and computes equilibrium strategies, humans increasingly function as \emph{supervisors}, establishing higher-level objectives and intervening only when AI reasoning encounters novel situations outside its training distribution and game-theoretic reasoning.

This role inversion carries profound implications for cybersecurity as a discipline. First, it fundamentally alters the economics of expertise: whereas traditional security required years of apprenticeship to develop intuition for attack surface enumeration, vulnerability chaining, and exploitation (knowledge that remained concentrated among security experts), AI systems now encode and operationalize this expertise at marginal cost approaching zero. The democratization effect is substantial: organizations previously excluded from sophisticated security assessment due to cost or talent scarcity can now access capabilities that exceeded best human expert performance just months prior. Second, the inversion reshapes the cognitive demands on security professionals. The transition from actor to supervisor does not diminish human importance but transforms its nature: supervisory competence requires meta-cognitive skills, including understanding AI capabilities and limitations, recognizing situations requiring human intervention, and maintaining strategic oversight without tactical immersion. This mirrors patterns observed in aviation and medicine, where automation paradoxically increases demands on human operators who must remain competent to intervene in systems they rarely control directly \cite{parasuraman2000model}. Third, the speed differential (AI operating 3,600$\times$ faster than humans \cite{aliasrobotics2025cai}) creates temporal asymmetries that challenge traditional security workflows. Incident response, vulnerability disclosure, and patch deployment processes designed around human timescales become bottlenecks in a regime where AI can enumerate attack surfaces faster than organizations can process findings. Finally, the emergence of strategic AI reasoning introduces a qualitatively new dynamic: when both offensive and defensive capabilities incorporate game-theoretic analysis, security becomes an algorithmic arms race where human strategic intuition may prove insufficient to supervise systems reasoning at superhuman speeds about superhuman strategies.

%Consistent with the progression and role inversion described, we define Cybersecurity Superintelligence as an aspirational AI capability exceeding the best human performance in both speed \emph{and} strategic reasoning. The neurosymbolic architecture demonstrated by G-CTR (symbolic game-theoretic analysis augmenting neural LLM inference) represents the most promising path toward this goal. This definition distinguishes our work from incremental automation: true security-specialized superintelligence requires not merely faster execution but qualitatively superior strategic reasoning that anticipates adversarial responses and adapts to novel scenarios.

The global impact of the research presented here validates the trajectory toward superintelligence in cybersecurity and is available for reproduction through the CAI framework~\cite{aliasrobotics2025cai,cai2025github}. CAI has grown to become the largest open-source AI security project, with 50,000+  users across 70 countries, 10,000+ URLs assessed, and a 1,500-member developer community. Regional adoption shows Europe (39\%), North America (27\%), and Asia (20\%) leading deployment. CAI enables organizations previously lacking specialized security expertise to access expert-level (and beyond) capabilities, a fundamental shift from security as an exclusive domain to security as accessible infrastructure.

Yet significant barriers remain to be solved before these superintelligence capabilities are fully realized and achieve widespread deployment. First, the economics of AI security agents present challenges: state-of-the-art LLMs cost approximately \$5,940 per billion tokens (equivalent to one month of continuous single agent operation), rendering sustained automated security economically unviable for most organizations. Recent work \cite{mayoral2025worldstopagent} demonstrates a solution through multi-model orchestration with entropy-based dynamic selection, achieving 98\% cost reduction (\$5,940→\$119 per billion tokens) while maintaining competitive performance. Second, there remains significant room for improving \emph{agency} in security solutions, the capacity for independent decision-making, strategic planning, and adaptive response. Common security tools predominantly occupy lower agency levels~\cite{mayoral2025cybersecurity}. Sustained progress requires continuous human-curated knowledge and data; without ongoing update and supervision, AI security agents risk performance drift and degradation, reinforcing the need not to rely solely on AI. %CAI delivers automation (supervised end-to-end execution with adaptive reasoning) while G-CTR adds strategic planning reducing ambiguity, collapsing the LLM's search space, suppressing hallucinations, and keeping the model tightly anchored to the most strategically relevant parts of the problem. Yet even these advances fall short of true autonomy which imply delegated decision-making capability. Real-world incident response expose fundamental gaps where autonomy, not merely sophisticated automation, remains aspirational. 
Even with CAI’s supervised automation and G-CTR’s strategic guidance, true autonomy—delegated decision-making—remains out of scope in real-world incident response.
%Critical challenges thus remain on the path to Cybersecurity Superintelligence.

With AI weaponized by nation-states, democratizing defensive capability via open-source frameworks is imperative. This transformation is empirically validated; our task is to steer it responsibly so that advances from human–AI collaboration to strategic AI strengthen defense rather than fuel offensive proliferation.
%As nation-states weaponize AI for offensive operations, democratizing defensive capabilities through open-source frameworks becomes imperative. The transformation of cybersecurity through AI is not theoretical but empirically validated. Our task is shaping this transformation responsibly, ensuring that the trajectory from human-AI collaboration to strategic AI guidance and supervision serves defensive democratization rather than offensive proliferation.

% \section*{Acknowledgements}
% We thank the CAI open-source community and security researchers who contribute to this research.

\section*{Author Contributions}
V.M.-V. conceived the study, led the overall research, designed and led the experiments, wrote the main manuscript, and served as the principal scientific lead of the three core contributions presented in this work (PentestGPT, CAI, and G-CTR). S.R. and M.P. contributed to the development of PentestGPT and CAI; S.R. additionally contributed to G-CTR. Both S.R. and M.P. contributed to the methodological design, scientific rigor, and validation of the study. E.G.-U., U.A.-C., J.A.R.-A., and M.d.M.d.T. contributed to the analysis and alignment of the state of the art and to the critical review of the perspectives and contributions of the work. M.S.-G. and F.B. contributed to CAI and G-CTR, including development, testing and experiments, validation, and scientific grounding. C.R.J.V.-C., V.T., A.G.-P., D.S.-P., Y.S., P.Z.-L., R.R.C.-Á., and P.M.-P. contributed to the testing and development of CAI. All authors reviewed and approved the final manuscript.

\section*{Declarations}
Funding: European Innovation Council (GA 101161136). Competing interests: None. Data/Code availability: https://github.com/aliasrobotics/cai (Dual MIT/Proprietary license).

\bibliography{csi-bibliography}

\newpage
\begin{appendices}
\section{Full Model Comparison}\label{appendix:full-model-comparison}

% Heatmap showing CTF results for ALL models (not just alias)
\begin{figure}[!h]
    \centering
    \resizebox{\linewidth}{!}{%
    \begin{tikzpicture}[
        remember picture,
    ]

    % ============================================================================
    % HEATMAP - CTF Scenario Comparison across ALL models
    % Ordered by number of solved scenarios (descending)
    % ============================================================================

    % Heatmap background - 19 models with space for rotated labels
    \fill[white, rounded corners=4pt] (-3.5, -13.75) rectangle (17.2, 0.3);
    \draw[graph_teal!30, line width=0.6pt, rounded corners=4pt] (-3.5, -13.75) rectangle (17.2, 0.3);

    % Define cell dimensions - 33 cells fitting in available width
    \pgfmathsetmacro{\cellw}{0.42}
    \pgfmathsetmacro{\cellh}{0.45}
    \pgfmathsetmacro{\startx}{0}
    \pgfmathsetmacro{\rowgap}{0.55}

    % Brand-coherent colors (matching Figure 2):
    \definecolor{anthropic_color}{RGB}{204, 119, 102}  % Anthropic coral/terracotta
    \definecolor{google_color}{RGB}{66, 133, 244}      % Google blue
    \definecolor{openai_color}{RGB}{100, 100, 100}     % Grey for OpenAI
    \definecolor{mistral_color}{RGB}{255, 140, 0}      % Mistral orange

    % Y-axis labels - All 19 models ordered by solve count
    % Row 0: claude opus 4.5 (27 solved)
    \node[font=\scriptsize\sffamily, text=graph_navy, anchor=east] at (-0.1, -0.275) {claude opus 4.5$^{\textcolor{cai_primary!50}{\text{11/25}}}$};
    % Row 1: alias2 (25 solved)
    \node[font=\scriptsize\sffamily, text=cai_primary, anchor=east] at (-0.1, -0.275-\rowgap) {\textbf{alias2}$^{\textcolor{cai_primary!50}{\text{01/26}}}$};
    % Row 2: claude opus 4.6 (24 solved)
    \node[font=\scriptsize\sffamily, text=graph_navy, anchor=east] at (-0.1, -0.275-2*\rowgap) {claude opus 4.6$^{\textcolor{cai_primary!50}{\text{02/26}}}$};
    % Row 3: gemini 3 pro (21 solved)
    \node[font=\scriptsize\sffamily, text=graph_navy, anchor=east] at (-0.1, -0.275-3*\rowgap) {gemini 3 pro$^{\textcolor{cai_primary!50}{\text{11/25}}}$};
    % Row 4: claude sonnet 4.5 (16 solved)
    \node[font=\scriptsize\sffamily, text=graph_navy, anchor=east] at (-0.1, -0.275-4*\rowgap) {claude sonnet 4.5$^{\textcolor{cai_primary!50}{\text{09/25}}}$};
    % Row 5: gpt 5.2 (16 solved)
    \node[font=\scriptsize\sffamily, text=graph_navy, anchor=east] at (-0.1, -0.275-5*\rowgap) {gpt 5.2$^{\textcolor{cai_primary!50}{\text{12/25}}}$};
    % Row 6: alias1 (14 solved)
    \node[font=\scriptsize\sffamily, text=cai_primary, anchor=east] at (-0.1, -0.275-6*\rowgap) {\textbf{alias1}$^{\textcolor{cai_primary!50}{\text{10/25}}}$};
    % Row 7: gpt 5.1 (13 solved)
    \node[font=\scriptsize\sffamily, text=graph_navy, anchor=east] at (-0.1, -0.275-7*\rowgap) {gpt 5.1$^{\textcolor{cai_primary!50}{\text{11/25}}}$};
    % Row 8: gpt 5 (11 solved)
    \node[font=\scriptsize\sffamily, text=graph_navy, anchor=east] at (-0.1, -0.275-8*\rowgap) {gpt 5$^{\textcolor{cai_primary!50}{\text{08/25}}}$};
    % Row 9: gemini 2.5 pro (7 solved)
    \node[font=\scriptsize\sffamily, text=graph_navy, anchor=east] at (-0.1, -0.275-9*\rowgap) {gemini 2.5 pro$^{\textcolor{cai_primary!50}{\text{03/25}}}$};
    % Row 10: claude 4 sonnet (7 solved)
    \node[font=\scriptsize\sffamily, text=graph_navy, anchor=east] at (-0.1, -0.275-10*\rowgap) {claude 4 sonnet$^{\textcolor{cai_primary!50}{\text{05/25}}}$};
    % Row 11: alias0 (6 solved)
    \node[font=\scriptsize\sffamily, text=cai_primary, anchor=east] at (-0.1, -0.275-11*\rowgap) {\textbf{alias0}$^{\textcolor{cai_primary!50}{\text{05/25}}}$};
    % Row 12: claude 3.5 sonnet (6 solved)
    \node[font=\scriptsize\sffamily, text=graph_navy, anchor=east] at (-0.1, -0.275-12*\rowgap) {claude 3.5 sonnet$^{\textcolor{cai_primary!50}{\text{06/24}}}$};
    % Row 13: claude 3 opus (6 solved)
    \node[font=\scriptsize\sffamily, text=graph_navy, anchor=east] at (-0.1, -0.275-13*\rowgap) {claude 3 opus$^{\textcolor{cai_primary!50}{\text{03/24}}}$};
    % Row 14: gpt-4o (5 solved)
    \node[font=\scriptsize\sffamily, text=graph_navy, anchor=east] at (-0.1, -0.275-14*\rowgap) {gpt-4o$^{\textcolor{cai_primary!50}{\text{05/24}}}$};
    % Row 15: mistral large 3 (4 solved)
    \node[font=\scriptsize\sffamily, text=graph_navy, anchor=east] at (-0.1, -0.275-15*\rowgap) {mistral large 3$^{\textcolor{cai_primary!50}{\text{12/25}}}$};
    % Row 16: mistral large 2.0 (4 solved)
    \node[font=\scriptsize\sffamily, text=graph_navy, anchor=east] at (-0.1, -0.275-16*\rowgap) {mistral large 2.0$^{\textcolor{cai_primary!50}{\text{07/24}}}$};
    % Row 17: gemini 1.5 pro (2 solved)
    \node[font=\scriptsize\sffamily, text=graph_navy, anchor=east] at (-0.1, -0.275-17*\rowgap) {gemini 1.5 pro$^{\textcolor{cai_primary!50}{\text{02/24}}}$};
    % Row 18: mistral large 2.1 (1 solved)
    \node[font=\scriptsize\sffamily, text=graph_navy, anchor=east] at (-0.1, -0.275-18*\rowgap) {mistral large 2.1$^{\textcolor{cai_primary!50}{\text{11/24}}}$};

    % Row 0: claude opus 4.5 (27 solved) - Anthropic coral
    \foreach \i/\status in {0/1,1/1,2/1,3/1,4/1,5/1,6/1,7/1,8/1,9/1,10/1,11/1,12/1,13/1,14/1,15/1,16/1,17/1,18/1,19/1,20/1,21/1,22/1,23/1,24/0,25/1,26/1,27/0,28/0,29/0,30/0,31/1,32/0} {
        \pgfmathsetmacro{\xpos}{\startx + \i*\cellw}
        \pgfmathsetmacro{\ypos}{-0.275}
        \ifnum\status=1
            \fill[anthropic_color!70, rounded corners=1pt] (\xpos, \ypos-\cellh/2) rectangle (\xpos+\cellw-0.04, \ypos+\cellh/2);
        \else
            \fill[graph_gray!40, rounded corners=1pt] (\xpos, \ypos-\cellh/2) rectangle (\xpos+\cellw-0.04, \ypos+\cellh/2);
        \fi
    }

    % Row 1: alias2 (25 solved) - teal color
    \foreach \i/\status in {0/1,1/1,2/1,3/1,4/1,5/1,6/1,7/1,8/1,9/1,10/1,11/1,12/0,13/1,14/1,15/0,16/1,17/1,18/1,19/1,20/1,21/1,22/1,23/1,24/0,25/1,26/0,27/0,28/1,29/0,30/0,31/1,32/0} {
        \pgfmathsetmacro{\xpos}{\startx + \i*\cellw}
        \pgfmathsetmacro{\ypos}{-0.275 - \rowgap}
        \ifnum\status=1
            \fill[graph_teal, rounded corners=1pt] (\xpos, \ypos-\cellh/2) rectangle (\xpos+\cellw-0.04, \ypos+\cellh/2);
        \else
            \fill[graph_gray!40, rounded corners=1pt] (\xpos, \ypos-\cellh/2) rectangle (\xpos+\cellw-0.04, \ypos+\cellh/2);
        \fi
    }

    % Row 2: claude opus 4.6 (24 solved) - Anthropic coral
    \foreach \i/\status in {0/1,1/1,2/1,3/1,4/1,5/1,6/1,7/1,8/1,9/1,10/1,11/0,12/1,13/1,14/1,15/0,16/1,17/1,18/0,19/1,20/0,21/1,22/1,23/1,24/0,25/1,26/1,27/1,28/0,29/0,30/0,31/1,32/0} {
        \pgfmathsetmacro{\xpos}{\startx + \i*\cellw}
        \pgfmathsetmacro{\ypos}{-0.275 - 2*\rowgap}
        \ifnum\status=1
            \fill[anthropic_color!70, rounded corners=1pt] (\xpos, \ypos-\cellh/2) rectangle (\xpos+\cellw-0.04, \ypos+\cellh/2);
        \else
            \fill[graph_gray!40, rounded corners=1pt] (\xpos, \ypos-\cellh/2) rectangle (\xpos+\cellw-0.04, \ypos+\cellh/2);
        \fi
    }

    % Row 3: gemini 3 pro (21 solved) - Google blue
    \foreach \i/\status in {0/1,1/1,2/1,3/1,4/1,5/1,6/1,7/1,8/1,9/1,10/1,11/1,12/1,13/1,14/1,15/0,16/1,17/0,18/0,19/1,20/0,21/1,22/0,23/1,24/0,25/0,26/0,27/0,28/0,29/0,30/0,31/0,32/0} {
        \pgfmathsetmacro{\xpos}{\startx + \i*\cellw}
        \pgfmathsetmacro{\ypos}{-0.275 - 3*\rowgap}
        \ifnum\status=1
            \fill[google_color!50, rounded corners=1pt] (\xpos, \ypos-\cellh/2) rectangle (\xpos+\cellw-0.04, \ypos+\cellh/2);
        \else
            \fill[graph_gray!40, rounded corners=1pt] (\xpos, \ypos-\cellh/2) rectangle (\xpos+\cellw-0.04, \ypos+\cellh/2);
        \fi
    }

    % Row 4: claude sonnet 4.5 (16 solved) - Anthropic coral
    \foreach \i/\status in {0/1,1/0,2/1,3/1,4/1,5/1,6/1,7/1,8/1,9/1,10/0,11/0,12/0,13/0,14/1,15/0,16/1,17/0,18/1,19/1,20/0,21/1,22/0,23/0,24/0,25/0,26/0,27/0,28/1,29/0,30/0,31/1,32/0} {
        \pgfmathsetmacro{\xpos}{\startx + \i*\cellw}
        \pgfmathsetmacro{\ypos}{-0.275 - 4*\rowgap}
        \ifnum\status=1
            \fill[anthropic_color!35, rounded corners=1pt] (\xpos, \ypos-\cellh/2) rectangle (\xpos+\cellw-0.04, \ypos+\cellh/2);
        \else
            \fill[graph_gray!40, rounded corners=1pt] (\xpos, \ypos-\cellh/2) rectangle (\xpos+\cellw-0.04, \ypos+\cellh/2);
        \fi
    }

    % Row 5: gpt 5.2 (16 solved) - OpenAI grey
    \foreach \i/\status in {0/1,1/1,2/1,3/1,4/1,5/1,6/0,7/1,8/1,9/1,10/0,11/0,12/0,13/1,14/1,15/0,16/1,17/0,18/0,19/1,20/0,21/1,22/0,23/0,24/0,25/0,26/0,27/1,28/0,29/0,30/0,31/1,32/0} {
        \pgfmathsetmacro{\xpos}{\startx + \i*\cellw}
        \pgfmathsetmacro{\ypos}{-0.275 - 5*\rowgap}
        \ifnum\status=1
            \fill[openai_color!50, rounded corners=1pt] (\xpos, \ypos-\cellh/2) rectangle (\xpos+\cellw-0.04, \ypos+\cellh/2);
        \else
            \fill[graph_gray!40, rounded corners=1pt] (\xpos, \ypos-\cellh/2) rectangle (\xpos+\cellw-0.04, \ypos+\cellh/2);
        \fi
    }

    % Row 6: alias1 (14 solved) - teal!60 color
    \foreach \i/\status in {0/1,1/1,2/1,3/1,4/1,5/1,6/0,7/0,8/0,9/0,10/0,11/1,12/1,13/1,14/1,15/0,16/1,17/1,18/0,19/0,20/0,21/1,22/0,23/0,24/0,25/0,26/0,27/0,28/0,29/0,30/0,31/0,32/0} {
        \pgfmathsetmacro{\xpos}{\startx + \i*\cellw}
        \pgfmathsetmacro{\ypos}{-0.275 - 6*\rowgap}
        \ifnum\status=1
            \fill[graph_teal!60, rounded corners=1pt] (\xpos, \ypos-\cellh/2) rectangle (\xpos+\cellw-0.04, \ypos+\cellh/2);
        \else
            \fill[graph_gray!40, rounded corners=1pt] (\xpos, \ypos-\cellh/2) rectangle (\xpos+\cellw-0.04, \ypos+\cellh/2);
        \fi
    }

    % Row 7: gpt 5.1 (13 solved) - OpenAI grey
    \foreach \i/\status in {0/1,1/0,2/1,3/1,4/1,5/1,6/0,7/1,8/0,9/1,10/0,11/0,12/0,13/1,14/1,15/0,16/1,17/0,18/0,19/1,20/0,21/1,22/0,23/0,24/0,25/0,26/0,27/1,28/0,29/0,30/0,31/0,32/0} {
        \pgfmathsetmacro{\xpos}{\startx + \i*\cellw}
        \pgfmathsetmacro{\ypos}{-0.275 - 7*\rowgap}
        \ifnum\status=1
            \fill[openai_color!50, rounded corners=1pt] (\xpos, \ypos-\cellh/2) rectangle (\xpos+\cellw-0.04, \ypos+\cellh/2);
        \else
            \fill[graph_gray!40, rounded corners=1pt] (\xpos, \ypos-\cellh/2) rectangle (\xpos+\cellw-0.04, \ypos+\cellh/2);
        \fi
    }

    % Row 8: gpt 5 (11 solved) - OpenAI grey
    \foreach \i/\status in {0/1,1/0,2/1,3/0,4/1,5/1,6/0,7/1,8/1,9/1,10/0,11/0,12/0,13/0,14/1,15/0,16/1,17/0,18/0,19/1,20/0,21/1,22/0,23/0,24/0,25/0,26/0,27/0,28/0,29/0,30/0,31/0,32/0} {
        \pgfmathsetmacro{\xpos}{\startx + \i*\cellw}
        \pgfmathsetmacro{\ypos}{-0.275 - 8*\rowgap}
        \ifnum\status=1
            \fill[openai_color!50, rounded corners=1pt] (\xpos, \ypos-\cellh/2) rectangle (\xpos+\cellw-0.04, \ypos+\cellh/2);
        \else
            \fill[graph_gray!40, rounded corners=1pt] (\xpos, \ypos-\cellh/2) rectangle (\xpos+\cellw-0.04, \ypos+\cellh/2);
        \fi
    }

    % Row 9: gemini 2.5 pro (7 solved) - Google blue
    \foreach \i/\status in {0/1,1/0,2/1,3/1,4/1,5/1,6/0,7/0,8/0,9/0,10/0,11/0,12/0,13/0,14/0,15/0,16/1,17/0,18/0,19/1,20/0,21/0,22/0,23/0,24/0,25/0,26/0,27/0,28/0,29/0,30/0,31/0,32/0} {
        \pgfmathsetmacro{\xpos}{\startx + \i*\cellw}
        \pgfmathsetmacro{\ypos}{-0.275 - 9*\rowgap}
        \ifnum\status=1
            \fill[google_color!50, rounded corners=1pt] (\xpos, \ypos-\cellh/2) rectangle (\xpos+\cellw-0.04, \ypos+\cellh/2);
        \else
            \fill[graph_gray!40, rounded corners=1pt] (\xpos, \ypos-\cellh/2) rectangle (\xpos+\cellw-0.04, \ypos+\cellh/2);
        \fi
    }

    % Row 10: claude 4 sonnet (7 solved) - Anthropic coral
    \foreach \i/\status in {0/1,1/0,2/1,3/0,4/1,5/1,6/0,7/0,8/1,9/0,10/0,11/0,12/0,13/0,14/0,15/0,16/1,17/0,18/0,19/0,20/0,21/1,22/0,23/0,24/0,25/0,26/0,27/0,28/0,29/0,30/0,31/0,32/0} {
        \pgfmathsetmacro{\xpos}{\startx + \i*\cellw}
        \pgfmathsetmacro{\ypos}{-0.275 - 10*\rowgap}
        \ifnum\status=1
            \fill[anthropic_color!35, rounded corners=1pt] (\xpos, \ypos-\cellh/2) rectangle (\xpos+\cellw-0.04, \ypos+\cellh/2);
        \else
            \fill[graph_gray!40, rounded corners=1pt] (\xpos, \ypos-\cellh/2) rectangle (\xpos+\cellw-0.04, \ypos+\cellh/2);
        \fi
    }

    % Row 11: alias0 (6 solved) - teal!30 color
    \foreach \i/\status in {0/1,1/0,2/0,3/1,4/1,5/1,6/0,7/0,8/0,9/0,10/0,11/0,12/0,13/1,14/0,15/0,16/0,17/0,18/0,19/0,20/0,21/0,22/0,23/0,24/0,25/0,26/0,27/0,28/0,29/0,30/0,31/0,32/0} {
        \pgfmathsetmacro{\xpos}{\startx + \i*\cellw}
        \pgfmathsetmacro{\ypos}{-0.275 - 11*\rowgap}
        \ifnum\status=1
            \fill[graph_teal!30, rounded corners=1pt] (\xpos, \ypos-\cellh/2) rectangle (\xpos+\cellw-0.04, \ypos+\cellh/2);
        \else
            \fill[graph_gray!40, rounded corners=1pt] (\xpos, \ypos-\cellh/2) rectangle (\xpos+\cellw-0.04, \ypos+\cellh/2);
        \fi
    }

    % Row 12: claude 3.5 sonnet (6 solved) - Anthropic coral (muted)
    \foreach \i/\status in {0/1,1/0,2/1,3/1,4/1,5/1,6/0,7/0,8/0,9/0,10/0,11/0,12/0,13/0,14/0,15/0,16/1,17/0,18/0,19/0,20/0,21/0,22/0,23/0,24/0,25/0,26/0,27/0,28/0,29/0,30/0,31/0,32/0} {
        \pgfmathsetmacro{\xpos}{\startx + \i*\cellw}
        \pgfmathsetmacro{\ypos}{-0.275 - 12*\rowgap}
        \ifnum\status=1
            \fill[anthropic_color!35, rounded corners=1pt] (\xpos, \ypos-\cellh/2) rectangle (\xpos+\cellw-0.04, \ypos+\cellh/2);
        \else
            \fill[graph_gray!40, rounded corners=1pt] (\xpos, \ypos-\cellh/2) rectangle (\xpos+\cellw-0.04, \ypos+\cellh/2);
        \fi
    }

    % Row 13: claude 3 opus (6 solved) - Anthropic coral (muted)
    \foreach \i/\status in {0/1,1/0,2/1,3/1,4/1,5/1,6/0,7/0,8/0,9/0,10/0,11/0,12/0,13/0,14/0,15/0,16/1,17/0,18/0,19/0,20/0,21/0,22/0,23/0,24/0,25/0,26/0,27/0,28/0,29/0,30/0,31/0,32/0} {
        \pgfmathsetmacro{\xpos}{\startx + \i*\cellw}
        \pgfmathsetmacro{\ypos}{-0.275 - 13*\rowgap}
        \ifnum\status=1
            \fill[anthropic_color!70, rounded corners=1pt] (\xpos, \ypos-\cellh/2) rectangle (\xpos+\cellw-0.04, \ypos+\cellh/2);
        \else
            \fill[graph_gray!40, rounded corners=1pt] (\xpos, \ypos-\cellh/2) rectangle (\xpos+\cellw-0.04, \ypos+\cellh/2);
        \fi
    }

    % Row 14: gpt-4o (5 solved) - OpenAI grey (muted)
    \foreach \i/\status in {0/1,1/0,2/1,3/0,4/1,5/1,6/0,7/0,8/0,9/0,10/0,11/0,12/0,13/0,14/0,15/0,16/0,17/0,18/0,19/1,20/0,21/0,22/0,23/0,24/0,25/0,26/0,27/0,28/0,29/0,30/0,31/0,32/0} {
        \pgfmathsetmacro{\xpos}{\startx + \i*\cellw}
        \pgfmathsetmacro{\ypos}{-0.275 - 14*\rowgap}
        \ifnum\status=1
            \fill[openai_color!50, rounded corners=1pt] (\xpos, \ypos-\cellh/2) rectangle (\xpos+\cellw-0.04, \ypos+\cellh/2);
        \else
            \fill[graph_gray!40, rounded corners=1pt] (\xpos, \ypos-\cellh/2) rectangle (\xpos+\cellw-0.04, \ypos+\cellh/2);
        \fi
    }

    % Row 15: mistral large 3 (4 solved) - Mistral orange (muted)
    \foreach \i/\status in {0/1,1/1,2/0,3/1,4/0,5/0,6/0,7/0,8/0,9/0,10/0,11/0,12/0,13/0,14/0,15/0,16/0,17/0,18/0,19/1,20/0,21/0,22/0,23/0,24/0,25/0,26/0,27/0,28/0,29/0,30/0,31/0,32/0} {
        \pgfmathsetmacro{\xpos}{\startx + \i*\cellw}
        \pgfmathsetmacro{\ypos}{-0.275 - 15*\rowgap}
        \ifnum\status=1
            \fill[mistral_color!50, rounded corners=1pt] (\xpos, \ypos-\cellh/2) rectangle (\xpos+\cellw-0.04, \ypos+\cellh/2);
        \else
            \fill[graph_gray!40, rounded corners=1pt] (\xpos, \ypos-\cellh/2) rectangle (\xpos+\cellw-0.04, \ypos+\cellh/2);
        \fi
    }

    % Row 16: mistral large 2.0 (4 solved) - Mistral orange (muted)
    \foreach \i/\status in {0/1,1/1,2/0,3/1,4/0,5/1,6/0,7/0,8/0,9/0,10/0,11/0,12/0,13/0,14/0,15/0,16/0,17/0,18/0,19/0,20/0,21/0,22/0,23/0,24/0,25/0,26/0,27/0,28/0,29/0,30/0,31/0,32/0} {
        \pgfmathsetmacro{\xpos}{\startx + \i*\cellw}
        \pgfmathsetmacro{\ypos}{-0.275 - 16*\rowgap}
        \ifnum\status=1
            \fill[mistral_color!50, rounded corners=1pt] (\xpos, \ypos-\cellh/2) rectangle (\xpos+\cellw-0.04, \ypos+\cellh/2);
        \else
            \fill[graph_gray!40, rounded corners=1pt] (\xpos, \ypos-\cellh/2) rectangle (\xpos+\cellw-0.04, \ypos+\cellh/2);
        \fi
    }

    % Row 17: gemini 1.5 pro (2 solved) - Google blue (muted)
    \foreach \i/\status in {0/1,1/0,2/1,3/0,4/0,5/0,6/0,7/0,8/0,9/0,10/0,11/0,12/0,13/0,14/0,15/0,16/0,17/0,18/0,19/0,20/0,21/0,22/0,23/0,24/0,25/0,26/0,27/0,28/0,29/0,30/0,31/0,32/0} {
        \pgfmathsetmacro{\xpos}{\startx + \i*\cellw}
        \pgfmathsetmacro{\ypos}{-0.275 - 17*\rowgap}
        \ifnum\status=1
            \fill[google_color!50, rounded corners=1pt] (\xpos, \ypos-\cellh/2) rectangle (\xpos+\cellw-0.04, \ypos+\cellh/2);
        \else
            \fill[graph_gray!40, rounded corners=1pt] (\xpos, \ypos-\cellh/2) rectangle (\xpos+\cellw-0.04, \ypos+\cellh/2);
        \fi
    }

    % Row 18: mistral large 2.1 (1 solved) - Mistral orange (muted)
    \foreach \i/\status in {0/0,1/0,2/0,3/0,4/1,5/0,6/0,7/0,8/0,9/0,10/0,11/0,12/0,13/0,14/0,15/0,16/0,17/0,18/0,19/0,20/0,21/0,22/0,23/0,24/0,25/0,26/0,27/0,28/0,29/0,30/0,31/0,32/0} {
        \pgfmathsetmacro{\xpos}{\startx + \i*\cellw}
        \pgfmathsetmacro{\ypos}{-0.275 - 18*\rowgap}
        \ifnum\status=1
            \fill[mistral_color!50, rounded corners=1pt] (\xpos, \ypos-\cellh/2) rectangle (\xpos+\cellw-0.04, \ypos+\cellh/2);
        \else
            \fill[graph_gray!40, rounded corners=1pt] (\xpos, \ypos-\cellh/2) rectangle (\xpos+\cellw-0.04, \ypos+\cellh/2);
        \fi
    }

    % X-axis rotated labels for each challenge - ordered by difficulty
    \foreach \i/\name in {
        0/Loot Stash \ding{110},
        1/Urgent \ding{110},
        2/Packed Away \ding{110},
        3/It Has Begun \ding{110},
        4/Dynastic \ding{110},
        5/Primary Knowledge \ding{110},
        6/Delulu \ding{110},
        7/Crushing \ding{110}\ding{110},
        8/Partial Tenacity \ding{110}\ding{110},
        9/Missing Bits \ding{110}\ding{110},
        10/Unbreakable \ding{110}\ding{110},
        11/Glacier Exchange \ding{110}\ding{110},
        12/Avatar \ding{110}\ding{110},
        13/Eval Me \ding{110}\ding{110},
        14/Back to the Past \ding{110}\ding{110},
        15/Data Siege \ding{110}\ding{110}\ding{110},
        16/RPGO \ding{110}\ding{110}\ding{110},
        17/Were Pickle Phreaks Revenge \ding{110}\ding{110}\ding{110},
        18/Lock Talk \ding{110}\ding{110}\ding{110},
        19/Skilift \ding{110}\ding{110}\ding{110},
        20/Failproof \ding{110}\ding{110}\ding{110},
        21/Permuted \ding{110}\ding{110}\ding{110}\ding{110},
        22/Flecks of Gold \ding{110}\ding{110}\ding{110}\ding{110},
        23/SLCG \ding{110}\ding{110}\ding{110}\ding{110},
        24/SOP \ding{110}\ding{110}\ding{110}\ding{110},
        25/Shuffled AES \ding{110}\ding{110}\ding{110}\ding{110},
        26/Noisy CRC \ding{110}\ding{110}\ding{110}\ding{110},
        27/Ezmaze \ding{110}\ding{110}\ding{110}\ding{110},
        28/Diffecient \ding{110}\ding{110}\ding{110}\ding{110},
        29/Noisier CRC \ding{110}\ding{110}\ding{110}\ding{110}\ding{110},
        30/Randsubware \ding{110}\ding{110}\ding{110}\ding{110}\ding{110},
        31/Robust CBC \ding{110}\ding{110}\ding{110}\ding{110}\ding{110},
        32/Just Another Pickle Jail \ding{110}\ding{110}\ding{110}\ding{110}\ding{110}
    } {
        \pgfmathsetmacro{\xpos}{\startx + \i*\cellw + \cellw/2 - 0.02}
        \node[font=\tiny\sffamily, text=graph_navy!70, rotate=45, anchor=east] at (\xpos, -0.275 - 19*\rowgap - 0.15) {\name};
    }

    % X-axis label
    \node[font=\small\sffamily, text=graph_navy!70] at (6.9, -13.45) {CTF Challenges in \texttt{CAIBench-Jeopardy CTFs (Cybench)}~\cite{sanzgomez2025cybersecurityaibenchmarkcaibench}, $pass@3$};

    % Legend - grouped by model family
    \node[font=\tiny\sffamily\bfseries, text=graph_navy] at (15.3, -0.1) {Legend};

    % alias group
    \node[font=\tiny\sffamily\bfseries, text=graph_teal, anchor=west] at (14.2, -0.4) {alias};
    \fill[graph_teal, rounded corners=1pt] (14.35, -0.55) rectangle (14.5, -0.68);
    \node[font=\tiny\sffamily, text=graph_navy, anchor=west] at (14.55, -0.615) {alias2 (25/33, 76\%)};
    \fill[graph_teal!60, rounded corners=1pt] (14.35, -0.78) rectangle (14.5, -0.91);
    \node[font=\tiny\sffamily, text=graph_navy, anchor=west] at (14.55, -0.845) {alias1 (14/33, 42\%)};
    \fill[graph_teal!30, rounded corners=1pt] (14.35, -1.01) rectangle (14.5, -1.14);
    \node[font=\tiny\sffamily, text=graph_navy, anchor=west] at (14.55, -1.075) {alias0 (6/33, 18\%)};

    % Claude group
    \node[font=\tiny\sffamily\bfseries, text=anthropic_color, anchor=west] at (14.2, -1.35) {Claude};
    \fill[anthropic_color!70, rounded corners=1pt] (14.35, -1.5) rectangle (14.5, -1.63);
    \node[font=\tiny\sffamily, text=graph_navy, anchor=west] at (14.55, -1.565) {opus 4.5 (27/33, 82\%)};
    \fill[anthropic_color!70, rounded corners=1pt] (14.35, -1.73) rectangle (14.5, -1.86);
    \node[font=\tiny\sffamily, text=graph_navy, anchor=west] at (14.55, -1.795) {opus 4.6 (24/33, 73\%)};
    \fill[anthropic_color!35, rounded corners=1pt] (14.35, -1.96) rectangle (14.5, -2.09);
    \node[font=\tiny\sffamily, text=graph_navy, anchor=west] at (14.55, -2.025) {sonnet 4.5 (16/33, 48\%)};
    \fill[anthropic_color!35, rounded corners=1pt] (14.35, -2.19) rectangle (14.5, -2.32);
    \node[font=\tiny\sffamily, text=graph_navy, anchor=west] at (14.55, -2.255) {4 sonnet (7/33, 21\%)};
    \fill[anthropic_color!35, rounded corners=1pt] (14.35, -2.42) rectangle (14.5, -2.55);
    \node[font=\tiny\sffamily, text=graph_navy, anchor=west] at (14.55, -2.485) {3.5 sonnet (6/33, 18\%)};
    \fill[anthropic_color!70, rounded corners=1pt] (14.35, -2.65) rectangle (14.5, -2.78);
    \node[font=\tiny\sffamily, text=graph_navy, anchor=west] at (14.55, -2.715) {3 opus (6/33, 18\%)};

    % Gemini group
    \node[font=\tiny\sffamily\bfseries, text=google_color, anchor=west] at (14.2, -2.99) {Gemini};
    \fill[google_color!50, rounded corners=1pt] (14.35, -3.14) rectangle (14.5, -3.27);
    \node[font=\tiny\sffamily, text=graph_navy, anchor=west] at (14.55, -3.205) {3 pro (21/33, 64\%)};
    \fill[google_color!50, rounded corners=1pt] (14.35, -3.37) rectangle (14.5, -3.5);
    \node[font=\tiny\sffamily, text=graph_navy, anchor=west] at (14.55, -3.435) {2.5 pro (7/33, 21\%)};
    \fill[google_color!50, rounded corners=1pt] (14.35, -3.6) rectangle (14.5, -3.73);
    \node[font=\tiny\sffamily, text=graph_navy, anchor=west] at (14.55, -3.665) {1.5 pro (2/33, 6\%)};

    % GPT group
    \node[font=\tiny\sffamily\bfseries, text=openai_color, anchor=west] at (14.2, -3.94) {GPT};
    \fill[openai_color!50, rounded corners=1pt] (14.35, -4.09) rectangle (14.5, -4.22);
    \node[font=\tiny\sffamily, text=graph_navy, anchor=west] at (14.55, -4.155) {5.2 (16/33, 48\%)};
    \fill[openai_color!50, rounded corners=1pt] (14.35, -4.32) rectangle (14.5, -4.45);
    \node[font=\tiny\sffamily, text=graph_navy, anchor=west] at (14.55, -4.385) {5.1 (13/33, 39\%)};
    \fill[openai_color!50, rounded corners=1pt] (14.35, -4.55) rectangle (14.5, -4.68);
    \node[font=\tiny\sffamily, text=graph_navy, anchor=west] at (14.55, -4.615) {5 (11/33, 33\%)};
    \fill[openai_color!50, rounded corners=1pt] (14.35, -4.78) rectangle (14.5, -4.91);
    \node[font=\tiny\sffamily, text=graph_navy, anchor=west] at (14.55, -4.845) {4o (5/33, 15\%)};

    % Mistral group
    \node[font=\tiny\sffamily\bfseries, text=mistral_color, anchor=west] at (14.2, -5.12) {Mistral};
    \fill[mistral_color!50, rounded corners=1pt] (14.35, -5.27) rectangle (14.5, -5.4);
    \node[font=\tiny\sffamily, text=graph_navy, anchor=west] at (14.55, -5.335) {large 3 (4/33, 12\%)};
    \fill[mistral_color!50, rounded corners=1pt] (14.35, -5.5) rectangle (14.5, -5.63);
    \node[font=\tiny\sffamily, text=graph_navy, anchor=west] at (14.55, -5.565) {large 2.0 (4/33, 12\%)};
    \fill[mistral_color!50, rounded corners=1pt] (14.35, -5.73) rectangle (14.5, -5.86);
    \node[font=\tiny\sffamily, text=graph_navy, anchor=west] at (14.55, -5.795) {large 2.1 (1/33, 3\%)};

    % Not solved
    \fill[graph_gray!40, rounded corners=1pt] (14.35, -6.07) rectangle (14.5, -6.2);
    \node[font=\tiny\sffamily, text=graph_navy, anchor=west] at (14.55, -6.135) {Not solved};

    \end{tikzpicture}%
    }% end resizebox
    \caption{Full comparison of all evaluated models on the \texttt{CAIBench-Jeopardy CTFs (Cybench)}~\cite{sanzgomez2025cybersecurityaibenchmarkcaibench} benchmark, complementing the temporal progression shown in Figure~\ref{fig:ctf-progression}. Models are ordered by number of challenges solved (descending), with the \texttt{alias} series highlighted in teal. Each experiment was run for a maximum of 300 agentic interactions, 245 minutes of computing time per challenge, a maximum of 40 USD per challenge on API model expenses, and with $pass@3$. Superscripts indicate model release dates (MM/YY).}
    \label{fig:heatmap-full}
\end{figure}

\end{appendices}

\end{document}